\documentclass[seceq]{ptptex}

\title{
Dynamical Theory of Generalized Matrices
}

\author{
Yoshiharu {\sc Kawamura}\footnote{E-mail:
haru@azusa.shinshu-u.ac.jp} %
}

\inst{
Department of Physics, Shinshu University, Matsumoto 390-8621, Japan
}

\recdate{
}

\abst{
We propose a generalization of spin algebra using multi-index objects, 
and a dynamical system analogous to matrix theory.
The system has a solution described by generalized spin representation matrices
and possesses a symmetry similar to the volume preserving diffeomorphism 
in the $p$-brane action.
}

\begin{document}

\maketitle

\section{Introduction}

Group theoretical analysis has been applied successfully 
to a wide range of physical systems
because they are often invariant under certain transformations,
and such symmetry transformations, in many cases, form a group.
Matrices can represent the action of such group elements.
Among physical quantities, 
the spin variables (on which representation matrices of ${\it su}(2)$ operate) 
have played important roles.
Relativistic particles are classified based on two kinds of spin variables
because the Lorentz algebra is essentially specified by ${\it su}(2) \times {\it su}(2)$.\cite{Wigner}
The spin variables and their extension appear in the non-commutative geometry, 
which is expected as a possible description of 
space-time at a fundamental level.\cite{Connes}
For example, the fuzzy 2-sphere is the non-commutative space 
whose coordinates are inherently representation matrices of the spin algebra.\cite{FuzzyS2}
This space is used in a matrix description of the spherical membrane.\cite{dWHN}\cite{sph-mem}
It also appears as a solution of matrix theory and matrix model
with Chern-Simons like term.\cite{Myers}\cite{Model-CS}
Models related to higher dimensional fuzzy spheres are examined in 
various contexts.\cite{GKP}\cite{FuzzyS2k}\cite{QHE}\cite{FuzzyCoset}\cite{S4}
Hence it is a challenging work to explore a generalization of spin algebra
and representation matrices in order to unveil as-yet-unknown systems.

Recently, a generalization of spin algebra has been proposed that is
based on three-index objects,
and the connection between triple commutation relations and uncertainty relations has
been discussed.\cite{YK-Gspin}
This algebra can be generalized using an $n$-fold product as multiplication
and an $n$-fold commutator among $n$-index objects, as will be discussed below.
Such $n$-index objects are called $n$-th power matrices, which are interpreted as
a generalization of matrix, and
a new type of mechanics has been proposed
based on them.\cite{YK-GHD}\cite{YK}
This type of mechanics can be regarded as a generalization of Heisenberg's matrix mechanics.
It has interesting properties, but
it is not yet clear whether it is applicable to real physical systems
and what physical meaning many-index objects possess.
To break an impasse on physical applications of generalized matrices, 
we shift our focus to another systems.
With the expectation that study on analogous systems of matrix theory and matrix model
will provide any information,
it is intriguing to explore symmetry properties
in dynamical systems of generalized matrices,
while keeping their classical counterparts in mind.
The candidate of classical analog is the system of $p$-brane.\cite{p-brane}

In this paper, we propose a generalization of spin algebra 
using $n$-th power matrices, and
a dynamical system analogous to matrix theory.
The system has a solution described by generalized spin representation matrices
and possesses a symmetry similar to the volume preserving diffeomorphism 
in the $p$-brane action.

This paper is organized as follows.
In the next section, we give a definition of generalized spin algebras, 
generalized spin representation matrices
and a variant of fuzzy sphere.
We study a generalization of matrix theory
based on generalized matrices in $\S$3.  
Section 4 is devoted to conclusions and discussion.
In Appendix A, we define $n$-th power matrices, an $n$-fold product,
an $n$-fold commutator and two kinds of trace operations.
As we will see from the definition of the $n$-fold product, 
we do not use the Einstein's summation rule 
that repeated indices are summed, to avoid confusion, throughout this paper.
In Appendix B, we study transformation properties of hermitian $n$-th power matrices.
We explain the classical analog of generalized spin algebra in Appendix C
and the framework of classical $p$-branes in Appendix D.

\section{Generalized spin algebra}

First we review the spin algebra ${\it su}(2)$.
This algebra is defined by
\begin{eqnarray}
[J^a, J^b]_{mn} = i \hbar \sum_c \varepsilon^{abc} (J^c)_{mn} ,
\label{spin}
\end{eqnarray}
where $J^a$, $(a = 1, 2, 3)$ are spin representation matrices, $\hbar$ is the reduced Planck constant,
and $\varepsilon^{abc}$ is the Levi-Civita symbol.
Matrices in the adjoint representation are the $3 \times 3$ matrices given by
\begin{eqnarray}
(J^a)_{mn} = -i \hbar \varepsilon^{amn} ,
\label{adj}
\end{eqnarray}
where each of the indices $m$ and $n$ runs from 1 to 3.

Let us generalize the spin algebra defined by (\ref{spin}) 
using hermitian $n$-th power matrices.
(See Appendix A for the definition of hermitian $n$-th power matrix.)
In analogy to (\ref{adj}), we define 
the $(n + 1) \times (n + 1) \times \cdots \times (n + 1)$ matrices 
that we consider as follows:
\begin{eqnarray}
(J^a)_{l_1 l_2 \cdots l_n} = -i \hbar_{(n)} \varepsilon^{a l_1 l_2 \cdots l_n} , ~~
(K^a)_{l_1 l_2 \cdots l_n} = \hbar_{(n)} |\varepsilon^{a l_1 l_2 \cdots l_n}|  ,
\label{adj-cubic}
\end{eqnarray}
where $\varepsilon^{a l_1 l_2 \cdots l_n}$ is the $(n + 1)$-dimensional Levi-Civita symbol,
each of the indices $a$ and $l_i$, ($i = 1, 2, \cdots, n$) runs from 1 to $n+1$
and $\hbar_{(n)}$ is a new physical constant.
Hereafter $\hbar_{(n)}$ sets to 1 for simplicity. We find that the generalized matrices $J^a$ and $K^a$ 
form the algebra:
\begin{eqnarray}
&~& [J^{a_1}, \cdots, J^{a_{n-2j}}, K^{a_{n-2j+1}}, \cdots, K^{a_n}] 
= (-1)^{j} i \sum_{a_{n+1}} \varepsilon^{a_1 a_2 \cdots a_{n+1}} J^{a_{n+1}} , 
\label{G-spin-alg-even1}\\
&~& [J^{a_1}, \cdots, J^{a_{n-2j-1}}, K^{a_{n-2j}}, \cdots, K^{a_n}] 
= (-1)^{j+1} i \sum_{a_{n+1}} \varepsilon^{a_1 a_2 \cdots a_{n+1}} K^{a_{n+1}} ,
\label{G-spin-alg-even2}
\end{eqnarray}
for an even integer $n$ and $j = 0, 1, \cdots, n/2$ and
\begin{eqnarray}
&~& [J^{a_1}, \cdots, J^{a_{n-2j}}, K^{a_{n-2j+1}}, \cdots, K^{a_n}] 
= (-1)^{j+1} i \sum_{a_{n+1}} \varepsilon^{a_1 a_2 \cdots a_{n+1}} K^{a_{n+1}} , 
\label{G-spin-alg-odd1}\\
&~& [J^{a_1}, \cdots, J^{a_{n-2j-1}}, K^{a_{n-2j}}, \cdots, K^{a_n}] 
= (-1)^{j+1} i \sum_{a_{n+1}} \varepsilon^{a_1 a_2 \cdots a_{n+1}} J^{a_{n+1}} ,
\label{G-spin-alg-odd2}
\end{eqnarray}
for an odd integer $n$ and $j = 0, 1, \cdots, (n - 1)/2$. 
Here the indices $l_i$ are omitted
and the $n$-fold commutator is defined by (\ref{commutator}).

There exists a subalgebra of the algebra defined 
by (\ref{G-spin-alg-even1}) and (\ref{G-spin-alg-even2}) 
whose elements are $G^{a} = (J^{1}, \cdots, J^{n+1})$ for an even integer $n$,
and the subalgebra is given by
\begin{eqnarray}
&~& [G^{a_1}, G^{a_2}, \cdots, G^{a_n}]_{l_1 l_2 \cdots l_n} 
 = i \sum_{a_{n+1}} \varepsilon^{a_1 a_2 \cdots a_n a_{n+1}} (G^{a_{n+1}})_{l_1 l_2 \cdots l_n} .
\label{G-spin-subalg-even}
\end{eqnarray}
In the same way,
there exists a subalgebra of the algebra defined 
by (\ref{G-spin-alg-odd1}) and (\ref{G-spin-alg-odd2})
whose elements consist of a suitable set of $J^{a_p}$ and $K^{a_q}$.
For example, the elements 
$G^{a} = (J^1, \cdots, J^{n}, K^{n+1})$ for an odd integer $n$ form the algebra given by
\begin{eqnarray}
&~& [G^{a_1}, G^{a_2}, \cdots, G^{a_n}]_{l_1 l_2 \cdots l_n} 
 = - i \sum_{a_{n+1}} \varepsilon^{a_1 a_2 \cdots a_n a_{n+1}} (G^{a_{n+1}})_{l_1 l_2 \cdots l_n} .
\label{G-spin-subalg-odd}
\end{eqnarray}
We refer to the algebra defined by (\ref{G-spin-subalg-even}) and (\ref{G-spin-subalg-odd})
as a $\lq$generalized spin algebra' and write down collectively,
\begin{eqnarray}
&~& [G^{a_1}, G^{a_2}, \cdots, G^{a_n}]_{l_1 l_2 \cdots l_n} 
 = (-1)^n i \sum_{a_{n+1}} \varepsilon^{a_1 a_2 \cdots a_n a_{n+1}} (G^{a_{n+1}})_{l_1 l_2 \cdots l_n} .
\label{G-spin-subalg}
\end{eqnarray}
We refer to the elements of generalized spin algebra
as $\lq$generalized spin representation matrices'. 
We explain the classical analog of generalized spin algebra 
using a generalization of Hamiltonian dynamics in Appendix C.
Filippov also proposed a generalization of Lie algebra using vectors in the
$n$-dimensional Euclidean space as elements and the vector product
as multiplication.\cite{Filippov}\footnote{See also Ref. \citen{Hoppe} for a generalization of 
Lie algebra.}
In the realization, the basis vectors form 
an analog of the generalized spin algebra (\ref{G-spin-subalg}).
Xiong obtained the algebra, which is essentially equivalent to the algebra (\ref{G-spin-subalg-even}),
using the $n$-th power matrices 
$(T_a)_{i_1 i_2 \cdots i_{2m}} \equiv \varepsilon_{a i_1 i_2 \cdots i_{2m}}$,
$(a = 1, \cdots, N=2m+1)$.\cite{Xiong}
We have generalized the construction to the case with an arbitrary integer $N$.

The elements $G^a$ satisfy the so-called $\lq$fundamental indentity':
\begin{eqnarray}
&~& [[G^{a_1}, \cdots, G^{a_n}], G^{a_{n+1}}, \cdots, G^{a_{2n-1}}]_{l_1 l_2 \cdots l_n}
\nonumber \\ 
&~& ~~~~~~~~ = \sum_{i=1}^{n} 
[G^{a_1}, \cdots, [G^{a_i}, G^{a_{n+1}}, \cdots, G^{a_{2n-1}}], \cdots, G^{a_n}]_{l_1 l_2 \cdots l_n} .
\label{fund-id}
\end{eqnarray}
This identity is regarded as an extension of the Jacobi identity.

For later convenience, we write down several formulae of 
generalized spin representation matrices $G^a$.
By using (\ref{G-spin-subalg}) and the relation 
$\sum_{a_{1}, \cdots, a_{n}} \varepsilon^{a a_1 \cdots a_{n}} \varepsilon^{b a_1 \cdots a_{n}} 
 = n! \delta_{a b}$,
we obtain the formula:
\begin{eqnarray} &~& (G^a)_{l_1 l_2 \cdots l_n} = \frac{-i}{n !} \sum_{a_{1}, a_2, \cdots, a_{n}} 
\varepsilon^{a a_1 a_2 \cdots a_{n}} [G^{a_1}, G^{a_2}, \cdots, G^{a_n}]_{l_1 l_2 \cdots l_n}
\nonumber \\
&~& ~~~~~~~~~~~~~~  = -i \sum_{a_{1}, a_2, \cdots, a_{n}}  
\varepsilon^{a a_1 a_2 \cdots a_{n}} (G^{a_1}G^{a_2} \cdots G^{a_n})_{l_1 l_2 \cdots l_n} .
\label{F-1}
\end{eqnarray}
{}From (\ref{F-1}), the following formulae are derived,
\begin{eqnarray}
&~& \sum_a \mbox{Tr}_{(2)}(G^a)^2 \equiv \sum_a
\sum_{l_{1}, \cdots, l_{n-1}, l_{n}} (G^a)_{l_1 \cdots l_{n-1} l_n}  (G^a)_{l_1 \cdots l_n l_{n-1}} 
\nonumber\\
&~& ~~~~~~ = -  \frac{1}{n !}  \sum_{a_{1}, \cdots, a_{n}} \sum_{l_{1}, \cdots, l_{n-1}, l_{n}}
 [G^{a_1}, \cdots, G^{a_n}]_{l_1 \cdots l_{n-1} l_n} 
 [G^{a_1}, \cdots, G^{a_n}]_{l_1 \cdots l_n l_{n-1}}
\nonumber\\
&~& ~~~~~~ = -  \frac{i}{n !} \sum_{a, a_{1}, \cdots, a_{n}} \sum_{l_{1}, \cdots, l_{n-1}, l_{n}}
\varepsilon^{a a_1 \cdots a_{n}} 
(G^a)_{l_1 \cdots l_{n-1} l_n} [G^{a_1}, \cdots, G^{a_n}]_{l_1 \cdots l_n l_{n-1}} 
\nonumber \\
&~& ~~~~~~ =  -  i \sum_{a, a_{1}, \cdots, a_{n}} \sum_{l_{1}, \cdots, l_{n-1}, l_{n}}
\varepsilon^{a a_1 \cdots a_{n}} 
(G^a)_{l_1 \cdots l_{n-1} l_n} (G^{a_1} \cdots G^{a_n})_{l_1 \cdots l_n l_{n-1}} ,
\label{F-2}
\end{eqnarray}
where $\mbox{Tr}_{(2)}$ is the second kind of trace operator defined by (\ref{Tr(2)B1B2}).


The coordinates $X^i$ of fuzzy 2-sphere is defined by the matrices $J^i$, ($i=1,2,3$)
in the spin $j$ representation as follows,\cite{FuzzyS2}
\begin{eqnarray}
(X^i)_{mn} = \frac{R}{\sqrt{j(j+1)}} (J^i)_{mn}  ,
\label{Xi-f2}
\end{eqnarray}
where $R$ is regarded as the radius of fuzzy 2-sphere.
The coordinates $X^i$ satisfy the relations:
\begin{eqnarray}
&~& [X^i, X^j]_{mn} = i \frac{R}{\sqrt{j(j+1)}} \sum_k \varepsilon^{ijk} (X^k)_{mn} ,
~~~~ \sum_i (X^i)^2_{mn} = R^2 \delta_{mn} ,
\label{Xi-f2k-rel}
\end{eqnarray}
where each of the indices $m$ and $n$ runs from 1 to $2j+1$.
In the same way, the coordinates $X^i$, ($i=1,2, \cdots, 2k+1$) 
of fuzzy $2k$-sphere are the (tensor products of) matrices 
which satisfy the relations:\cite{FuzzyS2k}
\begin{eqnarray}
&~& [X^{i_1}, X^{i_2}, \cdots, X^{i_{2k}}]_{mn} 
= i \zeta \sum_{i_{2k+1}} \varepsilon^{i_1 i_2 \cdots i_{2k} i_{2k+1}} (X^{i_{2k+1}})_{mn} ,
\label{Xi-f2-alg}\\
&~& \sum_i (X^i)^2_{mn} = R^2 \delta_{mn} ,
\label{Xi-fn-rel}
\end{eqnarray}
where $\zeta$ is a constant parameter.
These fuzzy spheres are typical examples of non-commutative space.

Now we propose a variant of fuzzy sphere based on hermitian $n$-th power matrices
$X^i$, ($i=1,2 \cdots, n+1$).
The variables $X^i$ are interpreted as the coordinates
which satisfy the relations:
\begin{eqnarray}
&~& [X^{i_1}, X^{i_2}, \cdots, X^{i_{n}}]_{l_1 l_2 \cdots l_n} 
= i \eta \sum_{i_{n+1}} \varepsilon^{i_1 i_2 \cdots i_{n} i_{n+1}} (X^{i_{n+1}})_{l_1 l_2 \cdots l_n} ,
\label{Xi-fn-alg-new}\\
&~& \sum_i \sum_{l_1, \cdots, l_{n-2}, k} 
 (X^i)_{l_1 \cdots l_{n-2} l_{n-1} k}(X^i)_{l_1 \cdots l_{n-2} k l_{n}} = R^2 \delta_{l_{n-1} l_n} ,
\label{Xi-fn-rel-new}
\end{eqnarray}
where $\eta$ is a constant parameter.
The variables describing this new kind of $n$-dimensional space 
is, in general, non-commutative and non-associative for the $n$-fold product (\ref{product}).
The above relations (\ref{Xi-fn-alg-new}) and (\ref{Xi-fn-rel-new})
are invariant under the rotation:
\begin{eqnarray}
&~& (X^i)_{l_1 l_2 \cdots l_n}   \to (X^i)'_{l_1 l_2 \cdots l_n} 
 = \sum_{j} O^i_j (X^j)_{l_1 l_2 \cdots l_n}  ,
\label{Rotation}
\end{eqnarray}
where $O^i_j$ are elements of $(n+1)$-dimensional orthogonal group.
We suppose that the infinitesimal rotation is generated by the transformation:
\begin{eqnarray}
&~& \delta (X^i)_{l_1 l_2 \cdots l_n} = \sum_j \theta^{ij} (X^j)_{l_1 l_2 \cdots l_n}
= i [\Theta_1, \cdots, \Theta_{n-1}, X^i]_{l_1 l_2 \cdots l_n} ,
\label{inf-rotation}
\end{eqnarray}
where $\theta^{ij}(=-\theta^{ji})$ are infinitesimal parameters
and $\Theta_k$, $(k=1, \cdots, n-1)$ are $\lq\lq$generators" of rotation.
If $\Theta_k$ are given by $\Theta_k = \sum_i \theta^{(k)}_i X^i$,
$\theta^{ij}$ is written 
\begin{eqnarray}
\theta^{ij} = - \eta \sum_{i_1, \cdots, i_{n-1}} 
\varepsilon^{i_1 \cdots i_{n-1} i j} \theta^{(1)}_{i_1} \cdots \theta^{(n-1)}_{i_{n-1}} ,
\label{theta}
\end{eqnarray}
where $\theta^{(k)}_i$ are infinitesimal parameters.

\section{Dynamical System of Generalized Matrices}



In this section, we study a generalization of matrix theory 
using hermitian $n$-th power matrices.
We write down the Lagrangian, the Hamiltonian, the equation of motion,
and a solution including generalized spin representation matrices, and 
study symmetry properties.

Let us study the system 
described by the Lagrangian:
\begin{eqnarray}
&~& L =  \frac{1}{2} \sum_i \sum_{l_{1}, l_{2}, \cdots, l_{n}}
 (D_0 X^i)_{l_1 l_2 \cdots l_n} (D_0 X^i)_{l_2 l_1 \cdots l_n}
\nonumber \\
&~& ~~~ +  \frac{\alpha}{n \cdot n !} \sum_{i_{1}, i_{2}, \cdots, i_{n}}
\sum_{l_{1}, l_{2}, \cdots, l_{n}}
 [X^{i_1}, X^{i_2}, \cdots, X^{i_n}]_{l_1 l_2 \cdots l_n}   [X^{i_1}, X^{i_2}, \cdots, X^{i_n}]_{l_2 l_1 \cdots l_n} 
\nonumber \\
&~& ~~~ - \beta \sum_{i}
\sum_{l_{1}, l_{2}, \cdots, l_{n}}
(X^i)_{l_1 l_2 \cdots l_n} (X^i)_{l_2 l_1 \cdots l_n}
\nonumber \\
&~& ~~~ - \gamma \frac{2i}{n+1} \sum_{i, i_{1}, i_{2}, \cdots, i_{n}}
\sum_{l_{1}, l_{2}, \cdots, l_{n}}
f^{i i_1 i_2 \cdots i_{n}} 
(X^i)_{l_1 l_2 \cdots l_n} (X^{i_1} X^{i_2} \cdots X^{i_n})_{l_2 l_1 \cdots l_n} ,
\label{L}
\end{eqnarray}
where $X^i = X^i(t)$, $(i = 1, 2, \cdots, N)$ are time-dependent hermitian $n$-th power matrices,  
$\alpha$, $\beta$ and $\gamma$ are real parameters 
and $f^{i i_1 i_2 \cdots i_{n}}$ are real antisymmetric parameters.
The covariant time-derivative $D_0$ is defined by
\begin{eqnarray}
&~& (D_0 X^i)_{l_1 l_2 \cdots l_n} \equiv \frac{d}{dt} (X^i(t))_{l_1 l_2 \cdots l_n}
 + i [A_1, \cdots, A_{n-1}, X^i(t)]_{l_1 l_2 \cdots l_n} 
\nonumber \\
&~& ~~~~~~~~~~~ = \frac{d}{dt} (X^i(t))_{l_1 l_2 \cdots l_n}
 + i \sum_{m_1, m_2, \cdots, m_n} {\cal{A}}(t)_{l_1 l_2 \cdots l_n}^{m_1 m_2 \cdots m_n}
 (X^i(t))_{m_1 m_2 \cdots m_n} ,
\label{D0X}
\end{eqnarray}
where $A_{k}$, $(k=1, \cdots, n-1)$ are hermitian $n$-th power matrices
and ${\cal {A}}(t)$ is the $\lq\lq$gauge field" of time.
The first and second lines in (\ref{D0X}) are similar to (\ref{D0X-p-2}) and (\ref{D0X-p}), respectively.
Let us require that the Leibniz rule on the covariant time-derivative hold 
for the $n$-fold commutator $[B_1(t), \cdots, B_n(t)]$ as follows,
\begin{eqnarray}
(D_0[B_1(t), \cdots, B_n(t)])_{l_1 l_2 \cdots l_n} 
 =  \sum_{l=1}^{n} [B_1(t), \cdots, D_0 B_l(t), \cdots, B_n(t)]_{l_1 l_2 \cdots l_n} .
\label{derivation}
\end{eqnarray}
The above requirement fulfills for an arbitrary matrix $(A_1)_{l_1 l_2}$
with respect to the usual commutator of $[B_1(t), B_2(t)]$, but
it does not necessarily hold for arbitrary $n$-th power matrices $(A_k)_{l_1 l_2 \cdots l_n}$
with respect to the $n$-fold commutator with $n \geq 3$.
We find that the Leibniz rule (\ref{derivation}) holds for arbitrary $B_l(t)$, $(l = 1, 2, \cdots, n)$
if $A_k$s are normal $n$-th power matrices and 
the antisymmetric object defined by
$A(t)_{l_1 l_2 \cdots l_n} \equiv (-1)^{n-1} \widetilde{(A_1 \cdots A_{n-1})}_{l_1 l_2 \cdots l_n}$ 
satisfies the cocycle condition:
\begin{eqnarray}
&~& (\delta A(t))_{m_0 m_1 \cdots m_n} \equiv 
\sum_{i=0}^{n} (-1)^{i} A(t)_{m_0 m_1 \cdots \hat{m}_i \cdots m_n} = 0 ,
\label{cocycle-A}
\end{eqnarray}
where the hatted indix $\hat{m}_i$ is omitted and see (\ref{widetilde}) for the definition of 
$\widetilde{(A_1 \cdots A_{n-1})}$.
Then the covariant time-derivative (\ref{D0X}) is written
\begin{eqnarray}
(D_0 X^i)_{l_1 l_2 \cdots l_n} \equiv \frac{d}{dt} (X^i(t))_{l_1 l_2 \cdots l_n}
 + i A(t)_{l_1 l_2 \cdots l_n} (X^i(t))_{l_1 l_2 \cdots l_n} .
\label{D0X2}
\end{eqnarray}

 In the case with $n=2$, $\alpha = 1$ and $\beta = \gamma = 0$,
the Lagrangian (\ref{L}) is reduced to the bosonic part of BFSS matrix theory
by setting $R = g l_s =1$.\cite{BFSS}
Here $R$ is the compactification radius, $g$ is the string coupling
constant and $l_s$ is the string length scale.
The term with $\gamma$ is regarded as a generalization of Myers term.
It is known that the Myers term appears in the presence of a background antisymmetric field.\cite{Myers}
The BFSS matrix theory describes the system of D0-branes and
there is a conjecture that it is a microscopic
description of M-theory in the light-front coordinates.\footnote{See \citen{Taylor}
for a comprehensive review of matrix theory.}
The Lagrangian of the matrix theory is derived 
through the dimensional reduction to $(0+1)$-dimensional one
from $(9+1)$-dimensional super Yang-Mills one.
The matrix theory is also interpreted as a regularization of supermembrane theory.\cite{dWHN}

There are several proposals for a $\lq\lq$discretization" or 
quantization of $p$-brane system.\cite{Hoppe,many-index,Mimic}.
Our realization using $n$-th power matrices can be one of them
because the first and second terms in (\ref{L}) are regarded 
as counterparts in (\ref{S-p-brane-lcg-NB}).
In our system with $n \geq 3$,
it is not clear whether such interesting physical implications exist
as in the BFSS matrix theory. 

The Hamiltonian is obtained by
\begin{eqnarray}
&~& H =  \frac{1}{2} \sum_i \sum_{l_{1}, l_{2}, \cdots, l_{n}}
 (\Pi^i)_{l_1 l_2 \cdots l_n} (\Pi^i)_{l_2 l_1 \cdots l_n}
\nonumber \\
&~& ~~~ -  \frac{\alpha}{n \cdot n !} \sum_{i_{1}, i_{2}, \cdots, i_{n}}
\sum_{l_{1}, l_{2}, \cdots, l_{n}}
 [X^{i_1}, X^{i_2}, \cdots, X^{i_n}]_{l_1 l_2 \cdots l_n} 
[X^{i_1}, X^{i_2}, \cdots, X^{i_n}]_{l_2 l_1 \cdots l_n} 
\nonumber \\
&~& ~~~ + \beta \sum_{i}
\sum_{l_{1}, l_{2}, \cdots, l_{n}}
(X^i)_{l_1 l_2 \cdots l_n} (X^i)_{l_2 l_1 \cdots l_n}
\nonumber \\
&~& ~~~ + \gamma \frac{2i}{n+1} \sum_{i, i_{1}, i_{2}, \cdots, i_{n}}
\sum_{l_{1}, l_{2}, \cdots, l_{n}}
f^{i i_1 i_2 \cdots i_{n}} 
(X^i)_{l_1 l_2 \cdots l_n} (X^{i_1} X^{i_2} \cdots X^{i_n})_{l_2 l_1 \cdots l_n} ,
\label{H}
\end{eqnarray}
where $\Pi^i$ is the canonical conjugate momentum of $X^i$.

The following equation of motion is derived from the Lagrangian (\ref{L}),
\begin{eqnarray}
&~&  (D_0^2 X^i)_{l_1 l_2 \cdots l_n}
  +  \frac{2\alpha}{n !}  \sum_{i_{1}, \cdots, i_{n-1}}   [X^{i_1}, \cdots, X^{i_{n-1}}, [X^{i_1}, \cdots, X^{i_{n-1}}, X^{i}]]_{l_1 l_2 \cdots l_n} 
\nonumber \\
&~& + 2\beta (X^i)_{l_1 l_2 \cdots l_n} + 2 i \gamma  \sum_{i_{1}, i_{2}, \cdots, i_{n}}
f^{i i_1 i_2 \cdots i_{n}} (X^{i_1} X^{i_2} \cdots X^{i_n})_{l_1 l_2 \cdots l_n} = 0 .
\label{Eq}
\end{eqnarray}

Now we consider a case that 
$f^{a_1 a_2 \cdots a_{n+1}} = \varepsilon^{a_1 a_2 \cdots a_{n+1}}$, $(a_k, k = 1, 2, \cdots, n+1)$, 
and others components of $f^{i i_1 i_2 \cdots i_{n}}$ vanish. 
In this case, we find a non-trivial solution:
\begin{eqnarray}
 (X^a)_{l_1 l_2 \cdots l_n} = \xi (G^a)_{l_1 l_2 \cdots l_n} ,
~~~ (X^q)_{l_1 l_2 \cdots l_n} = 0 , 
~~~ {\cal{A}}(t)_{l_1 l_2 \cdots l_n}^{m_1 m_2 \cdots m_n} = 0 ,
\label{Sol}
\end{eqnarray}
where $G^a$, $(a = 1, 2, \cdots, n+1)$ are generalized spin representation matrices  
and $q$ runs other indices $(q = n+2, \cdots, N)$.
The parameter $\xi$ depends on $\alpha$, $\beta$ and $\gamma$
such as
\begin{eqnarray}
&~&  \xi = \left(\frac{\gamma \pm \sqrt{\gamma^2 - 4 \alpha \beta}}{2\alpha}\right)^{\frac{1}{n-1}} .
\label{xi}
\end{eqnarray}
This solution is interpreted as the counterpart of $n$-brane solution in the BFSS matrix theory.
For simplicity, we consider the case with $\beta =0$ and $\alpha = \gamma =1$.
Then we have the solution with $\xi = 1$:
\begin{eqnarray}
&~&  (X^a)_{l_1 l_2 \cdots l_n} = (x^a + v^a t) \delta_{l_1 l_2 \cdots l_n}
+  (G^a)_{l_1 l_2 \cdots l_n} ,
\nonumber \\
&~&  (X^q)_{l_1 l_2 \cdots l_n} = (x^q + v^q t) \delta_{l_1 l_2 \cdots l_n} ,
~~~ {\cal{A}}(t)_{l_1 l_2 \cdots l_n}^{m_1 m_2 \cdots m_n} = 0 ,
\label{Sol2}
\end{eqnarray}
where $x^a$, $v^a$, $x^q$ and $v^q$ are constants and 
$\delta_{l_1 l_2 \cdots l_n}=\delta_{l_1 l_2} \cdots \delta_{l_{n-1} l_n}$. The Hamiltonian takes a negative value for this solution,
\begin{eqnarray}
&~&  H = \frac{1-n}{n(n+1)} \sum_a \sum_{l_1, l_2, \cdots, l_n}
(G^a)_{l_1 l_2 \cdots l_n}(G^a)_{l_2 l_1 \cdots l_n} .
\label{<H>}
\end{eqnarray}
Hence the energy eigenvalue of this vacuum is lower than that of trivial solution, i.e.,
 $X^i = {\cal{A}}(t)= 0$.

Next we study symmetry properties of the above system.
(See Appendix B for a discussion on transformation properties of hermitian $n$-th power matrices.)
The system for $n=2$ is invariant under the time-dependent unitary transformation:
\begin{eqnarray}
&~& (X^i(t))_{l_1 l_2}   \to (X^i(t))'_{l_1 l_2} 
= \sum_{m_1, m_2} U(t)_{l_1 m_1} (X^i(t))_{m_1 m_2} U(t)_{m_2 l_2}^{\dagger} ,
\label{U(t)-transf-X}\\
&~& (A_1(t))_{l_1 l_2}  \to (A'_1(t))_{l_1 l_2} 
= \sum_{m_1, m_2} U(t)_{l_1 m_1} (A_1(t))_{m_1 m_2} U(t)_{m_2 l_2}^{\dagger}
\nonumber \\
&~& ~~~~~~~~~~~~~~~~~~~~~~~~~~~~~~~~~~~ 
 + i \sum_{m} \frac{d}{dt} U(t)_{l_1 m} \cdot U(t)_{m l_2}^{\dagger} ,
\label{U(t)-transf-A}
\end{eqnarray}
where $U(t)_{l m}$ is an arbitrary unitary matrix. 
The infinitesimal transformations are given by
\begin{eqnarray}
&~& \delta (X^i(t))_{l_1 l_2} = i[\Lambda(t), X^i(t)]_{l_1 l_2}   ,
\label{inf-U(t)-transf-X}\\
&~& \delta (A_1(t))_{l_1 l_2}  = - \frac{d}{dt}\Lambda(t)_{l_1 l_2} + i[\Lambda(t), A_1(t)]_{l_1 l_2} ,
\label{inf-U(t)-transf-A}
\end{eqnarray}
where $\Lambda(t)$ is the hermitian matrix which is related to $U(t)$ as $U(t) = \exp(i\Lambda(t))$.
The transformations (\ref{U(t)-transf-X})
and (\ref{inf-U(t)-transf-X}) are rewritten 
\begin{eqnarray}
&~& (X^i(t))_{l_1 l_2}   \to (X^i(t))'_{l_1 l_2} 
= \sum_{m_1, m_2} R(t)_{l_1 l_2}^{m_1 m_2} (X^i(t))_{m_1 m_2} ,
\label{U(t)-transf-X-R}\\
&~& \delta (X^i(t))_{l_1 l_2}
= i \sum_{m_1, m_2} \lambda(t)_{l_1 l_2}^{m_1 m_2} (X^i(t))_{m_1 m_2} ,
\label{inf-U(t)-transf-X-r}
\end{eqnarray}
respectively.
Here $R(t)$ and $\lambda(t)$ are $\lq\lq$transformation matrices"
given by $R(t)_{l_1 l_2}^{m_1 m_2} = U(t)_{l_1 m_1} U(t)_{l_2 m_2}^{*}$
and  $\lambda(t)_{l_1 l_2}^{m_1 m_2} = \Lambda(t)_{l_1 m_1} \delta_{l_2 m_2} 
 - \Lambda(t)_{m_2 l_2} \delta_{l_1 m_1}$.
They are related to as $R(t) = \exp(i\lambda(t))$.
(Note that we here use a different notation on the transformation matrix $(\lambda(t))$
from that $({r^{(\Lambda)}})$ in Appendix B.)
In terms of $R(t)$ and $\lambda(t)$, the finite and infinitesimal transformations 
of ${\cal{A}}(t)$ are given by
\begin{eqnarray}
&~&  {\cal{A}}(t)_{l_1 l_2}^{m_1 m_2} \to {\cal{A}}'(t)_{l_1 l_2}^{m_1 m_2} =
\sum_{n_1, n_2} \sum_{k_1, k_2} R(t)_{l_1 l_2}^{n_1 n_2} {\cal{A}}(t)_{n_1 n_2}^{k_1 k_2} 
 {R(t)^{-1}}_{k_1 k_2}^{m_1 m_2} 
\nonumber \\
&~& ~~~~~~~~~~~~~~~~~~~~  + \sum_{n_1, n_2} \frac{d}{dt} R(t)_{l_1 l_2}^{n_1 n_2} 
 \cdot {R(t)^{-1}}_{n_1 n_2}^{m_1 m_2} ,
\label{U(t)-transf-calA}\\
&~& \delta {\cal{A}}(t)_{l_1 l_2}^{m_1 m_2}  
 = - \frac{d}{dt}\lambda(t)_{l_1 l_2}^{m_1 m_2} 
 - i \sum_{n_1, n_2} {\cal{A}}(t)_{l_1 l_2}^{n_1 n_2} \lambda(t)_{n_1 n_2}^{m_1 m_2} 
\nonumber \\
&~& ~~~~~~~~~~~~~~~~~~~~ + i \sum_{n_1, n_2} \lambda(t)_{l_1 l_2}^{n_1 n_2} 
{\cal{A}}(t)_{n_1 n_2}^{m_1 m_2} ,
\label{inf-U(t)-transf-calA}
\end{eqnarray}
respectively. 
Here ${\cal{A}}(t)_{l_1 l_2}^{m_1 m_2} = {A_1(t)}_{l_1 m_1} \delta_{l_2 m_2} 
 - {A_1(t)}_{m_2 l_2} \delta_{l_1 m_1}$
and ${R(t)^{-1}}$ is the inverse transformation matrix of $R(t)$, which satisfy the relations:
\begin{eqnarray}
\sum_{n_1, n_2} R(t)_{l_1 l_2}^{n_1 n_2} {R(t)^{-1}}_{n_1 n_2}^{m_1 m_2} 
= \sum_{n_1, n_2} {R(t)^{-1}}_{l_1 l_2}^{n_1 n_2} R(t)_{n_1 n_2}^{m_1 m_2}
= \delta_{l_1 m_1}\delta_{l_2 m_2} .
\label{R-1}
\end{eqnarray}

Now let us study an extension 
of the unitary transformation of $X^i(t)$ and ${\cal{A}}(t)$
for the case with $n \geq 3$.
First we consider the infinitesimal transformation of $X^i(t)$
generated by a set of $\lq\lq$generators" $\Lambda_k$, $(k = 1, 2, \cdots, n-1)$ 
through the $n$-fold commutator such that
\begin{eqnarray}
&~& \delta (X^i(t))_{l_1 l_2 \cdots l_n} 
= i[\Lambda_1, \cdots, \Lambda_{n-1}, X^i(t)]_{l_1 l_2 \cdots l_n}  
\nonumber \\ 
&~& ~~~~~~~~~~~~~~~~~~~ = i \sum_{m_1, m_2, \cdots, m_n} 
\lambda(t)_{l_1 l_2 \cdots l_n}^{m_1 m_2 \cdots m_n} 
(X^i(t))_{m_1 m_2 \cdots m_n} .
\label{inf-U(t)-transf-X-n}
\end{eqnarray}
The expression (\ref{inf-U(t)-transf-X-n})
is similar to (\ref{VPD-X-NB}) and (\ref{VPD-X}).
Under the transformation (\ref{inf-U(t)-transf-X-n}),
the covariant time-derivative $D_0X^i$ transforms covariantly,
\begin{eqnarray}
&~& \delta (D_0 X^i)_{l_1 l_2 \cdots l_n} 
= i \sum_{m_1, m_2, \cdots, m_n} \lambda(t)_{l_1 l_2 \cdots l_n}^{m_1 m_2 \cdots m_n} 
(D_0 X^i)_{m_1 m_2 \cdots m_n} ,
\label{inf-U(t)-transf-D0X-n}
\end{eqnarray}
if ${\cal{A}}(t)$ transforms simultaneously as
\begin{eqnarray}
&~& \delta {\cal{A}}(t)_{l_1 l_2 \cdots l_n}^{m_1 m_2 \cdots m_n}  
 = - \frac{d}{dt}\lambda(t)_{l_1 l_2 \cdots l_n}^{m_1 m_2 \cdots m_n} 
 - i \sum_{k_1, k_2, \cdots, k_n} 
 {\cal{A}}(t)_{l_1 l_2 \cdots l_n}^{k_1 k_2 \cdots k_n} 
 \lambda(t)_{k_1 k_2 \cdots k_n}^{m_1 m_2 \cdots m_n} 
\nonumber \\
&~& ~~~~~~~~~~~~~~~~~~~~~~ 
+ i \sum_{k_1, k_2, \cdots, k_n} 
 \lambda(t)_{l_1 l_2 \cdots l_n}^{k_1 k_2 \cdots k_n} 
 {\cal{A}}(t)_{k_1 k_2 \cdots k_n}^{m_1 m_2 \cdots m_n} .
\label{inf-U(t)-transf-calA-n}
\end{eqnarray}
The expression (\ref{inf-U(t)-transf-calA-n}) is similar to (\ref{VPD-ua}).
We can show that the first and third terms in (\ref{L}) are invariant under the above infinitesimal
transformations (\ref{inf-U(t)-transf-X-n}) and (\ref{inf-U(t)-transf-calA-n}).
On the other hand, the second and fourth terms in (\ref{L}) are not 
necessarily invariant because
the $n$-fold commutator $[X^{i_1}, X^{i_2}, \cdots, X^{i_n}]$ transforms as
\begin{eqnarray}
&~& \delta [X^{i_1}, X^{i_2}, \cdots, X^{i_n}]_{l_1 l_2 \cdots l_n} 
= \sum_{k = 1}^{n} [X^{i_1}, \cdots, \delta(X^{i_k}), \cdots, X^{i_n}]_{l_1 l_2 \cdots l_n}
\nonumber \\
&~& ~~~ = i \sum_p \sum_{(j_1, \cdots, j_n)} \sum_k \mbox{sgn}(P) 
(X^{j_1})_{l_1 \cdots l_{n-1} k}  
\nonumber \\
&~& ~~~~ \cdots \sum_{m_1, m_2, \cdots, m_n} 
\lambda(t)_{l_1 \cdots l_{n-p} k l_{n+2-p} \cdots l_n}^{m_1 m_2 \cdots m_n} 
(X^{j_p})_{m_1 m_2 \cdots m_n} 
 \cdots (X^{j_n})_{k l_2 \cdots l_n} 
\label{inf-U(t)-transf-commutator-n}
\end{eqnarray}
under the transformation (\ref{inf-U(t)-transf-X-n}), and 
the transformation (\ref{inf-U(t)-transf-commutator-n}) do not always turn out the covariant form.
If the $[X^{i_1}, X^{i_2}, \cdots, X^{i_n}]$ transforms covariantly, i.e., 
\begin{eqnarray}
&~& \delta [X^{i_1}, X^{i_2}, \cdots, X^{i_n}]_{l_1 l_2 \cdots l_n} 
\nonumber \\
&~& ~~~~~~~~~~~~ = i \sum_{m_1, m_2, \cdots, m_n} \lambda(t)_{l_1 l_2 \cdots l_n}^{m_1 m_2 \cdots m_n} 
[X^{i_1}, X^{i_2}, \cdots, X^{i_n}]_{m_1 m_2 \cdots m_n} ,
\label{if-transf-comm}
\end{eqnarray}
our whole system possesses the local symmetry.

We discuss the case that the covariant time-derivative is given by (\ref{D0X2}). 
In this case, $(D_0 X^i)$ is invariant under
the transformation of $X^i(t)$ and $A(t)$ given by
\begin{eqnarray}
&~& \delta (X^i(t))_{l_1 l_2 \cdots l_n}
= i[\Lambda_1, \cdots, \Lambda_{n-1}, X^i(t)]_{l_1 l_2 \cdots l_n}  
\nonumber \\ 
&~& ~~~~~~~~~~~~~~~~~~~  = i \Lambda(t)_{l_1 l_2 \cdots l_n} (X^i(t))_{l_1 l_2 \cdots l_n} ,
\label{deltaX}\\
&~& \delta A(t)_{l_1 l_2 \cdots l_n} 
= - \frac{d}{dt} \Lambda(t)_{l_1 l_2 \cdots l_n}  ,
\label{deltaA}
\end{eqnarray}
where $\Lambda_k$, $(k = 1, \cdots, n-1)$ are real normal $n$-power matrices
and $\Lambda(t)$ is a real antisymmetric object defined by
\begin{eqnarray}
&~& \Lambda(t)_{l_1 l_2 \cdots l_n} 
\equiv (-1)^{n-1} \widetilde{(\Lambda_1 \cdots \Lambda_{n-1})}_{l_1 l_2 \cdots l_n}  .
\label{widetilde-Lambda}
\end{eqnarray}
The $\Lambda(t)$ should have the property:
\begin{eqnarray}
(\delta \Lambda(t))_{m_0 m_1 \cdots m_n} \equiv 
\sum_{i=0}^{n} (-1)^{i} \Lambda(t)_{m_0 m_1 \cdots \hat{m}_i \cdots m_n} = 0 ,
\label{cocycle}
\end{eqnarray} 
because $A(t) + \delta A(t)$ should do from the requirement of the Leibniz rule (\ref{derivation}).
When the antisymmetric objects $A(t)$ and $\Lambda(t)$ are treated as $n$-th power matrices,
the transformation (\ref{deltaA}) is rewritten 
\begin{eqnarray}
&~& \delta A(t)_{l_1 l_2 \cdots l_n} 
= - \frac{d}{dt} \Lambda(t)_{l_1 l_2 \cdots l_n}  
- i[A_1, \cdots, A_{n-1}, \Lambda(t)]_{l_1 l_2 \cdots l_n}
\nonumber \\
&~& ~~~~~~~~~~~~~~~~~~~~~ 
+ i[\Lambda_1, \cdots, \Lambda_{n-1}, A(t)]_{l_1 l_2 \cdots l_n} .
\label{deltaA-2}
\end{eqnarray}
Note that the last two terms in (\ref{deltaA-2}) are canceled out.
The above expression (\ref{deltaA-2}) is similar to (\ref{VPD-ua-NB}).
The finite versions of the transformations (\ref{deltaX}) and (\ref{deltaA}) are given by
\begin{eqnarray}
&~& (X^i(t))_{l_1 l_2 \cdots l_n}   \to (X^i(t))'_{l_1 l_2 \cdots l_n} 
=  e^{i\Lambda(t)_{l_1 l_2 \cdots l_n}} (X^i(t))_{l_1 l_2 \cdots l_n}  ,
\label{G-transf-X}\\
&~& A(t)_{l_1 l_2 \cdots l_n} \to A(t)'_{l_1 l_2 \cdots l_n} 
= A(t)_{l_1 l_2 \cdots l_n} - \frac{d}{dt}\Lambda(t)_{l_1 l_2 \cdots l_n} ,
\label{G-transf-A}
\end{eqnarray}
respectively.
It is easy to see that the Lagrangian (\ref{L}) is invariant 
under the transformations (\ref{deltaX}) and (\ref{deltaA})
or (\ref{G-transf-X}) and (\ref{G-transf-A}).
If $A(t)$ is a coboundary, i.e., $A(t) = \delta \Omega(t)$,
there is the extra symmetry that $A(t)$ is invariant under the
transformation:
\begin{eqnarray}
&~& \Omega(t)_{m_1 m_2 \cdots m_{n-1}} \to \Omega'(t)_{m_1 m_2 \cdots m_{n-1}}
\nonumber \\
&~& ~~~~~~~~~~~~~~~~~~~~~ = \Omega(t)_{m_1 m_2 \cdots m_{n-1}} 
  + (\delta \Theta(t))_{m_1 m_2 \cdots m_{n-1}} ,
\label{Omega-transf}
\end{eqnarray} 
where $\Theta(t)$ is an $(n-2)$-th rank of antisymmetric object.

Next we discuss a generalization of unitary transformations (\ref{U(t)-transf-X-R}) 
and (\ref{U(t)-transf-calA}), which are given by
\begin{eqnarray}
&~& (X^i(t))_{l_1 l_2 \cdots l_n}   \to (X^i(t))'_{l_1 l_2 \cdots l_n} 
\nonumber \\
&~& ~~~~~~~~~~~~~~~~~  = \sum_{m_1, m_2, \cdots, m_n} R(t)_{l_1 l_2 \cdots l_n}^{m_1 m_2 \cdots m_n} 
 (X^i(t))_{m_1 m_2 \cdots m_n}  ,
\label{G-transf-X-n}\\
&~&  {\cal{A}}(t)_{l_1 l_2 \cdots l_n}^{m_1 m_2 \cdots m_n} 
 \to {\cal{A}}'(t)_{l_1 l_2 \cdots l_n}^{m_1 m_2 \cdots m_n} 
\nonumber \\
&~& ~~~~~~~~~~~~~~~~~
= \sum_{n_1, n_2, \cdots, n_n} \sum_{k_1, k_2, \cdots, k_n} 
R(t)_{l_1 l_2 \cdots l_n}^{n_1 n_2 \cdots n_n} {\cal{A}}(t)_{n_1 n_2 \cdots n_n}^{k_1 k_2 \cdots k_n} 
 {R(t)^{-1}}_{k_1 k_2 \cdots k_n}^{m_1 m_2 \cdots m_n} 
\nonumber \\
&~& ~~~~~~~~~~~~~~~~~~~~~~  + \sum_{n_1, n_2, \cdots, l_n} 
\frac{d}{dt} R(t)_{l_1 l_2 \cdots l_n}^{n_1 n_2 \cdots n_n} \cdot
{R(t)^{-1}}_{n_1 n_2 \cdots n_n}^{m_1 m_2 \cdots m_n}  ,
\label{MoreG-tranf-calA-n}
\end{eqnarray} where $R(t)$ is a $\lq\lq$transformation matrix" and  $R(t)^{-1}$ is the inverse one.
In the case that $R(t)$ is factorized into a product of matrices such that 
\begin{eqnarray}
R_{l_1 l_2 \cdots l_n}^{m_1 m_2 \cdots m_n} 
= V_{l_1}^{m_1} V_{l_2}^{m_2} \cdots V_{l_n}^{m_n} ,
\label{R-n}
\end{eqnarray}
the first and third terms in (\ref{L}) are invariant under the transformation
that $V_l^m$ is an orthogonal matrix $O_l^m$.
We find that the Lagrangian (\ref{L}) is invariant under the discrete transformation such that 
$V_l^m = \delta_{l}^{\sigma(m)}$.
Here $\sigma(m)$ stands for the permutation among indices.

We have studied transformation properties of the system described
by the Lagrangian (\ref{L}).
There is a similarity between generalizations of unitary transformation
in the dynamical system of generalized matrices and 
the volume preserving diffeomorphism in the classical system of $p$-branes.
It is an important subject to explore the relationship between two systems
and make clear whether our theory describes microscopic physics
of $p$-brane like extended objects.


Finally we comment on other similar systems.

(i) Supersymmetric model\\
We write down the supersymmetric version of the Lagrangian (\ref{L})
with $\alpha = 1$ and $\beta = \gamma = 0$:
\begin{eqnarray}
&~& L =  \frac{1}{2} \sum_i \sum_{l_{1}, l_{2}, \cdots, l_{n}}
 (D_0 X^i)_{l_1 l_2 \cdots l_n} (D_0 X^i)_{l_2 l_1 \cdots l_n}
\nonumber \\
&~&  +  \frac{1}{n \cdot n !} \sum_{i_{1}, i_{2}, \cdots, i_{n}}
\sum_{l_{1}, l_{2}, \cdots, l_{n}}
 [X^{i_1}, X^{i_2}, \cdots, X^{i_n}]_{l_1 l_2 \cdots l_n} 
[X^{i_1}, X^{i_2}, \cdots, X^{i_n}]_{l_2 l_1 \cdots l_n} 
\nonumber \\
&~&  + \frac{i}{2} \sum_{l_{1}, l_{2}, \cdots, l_{n}}
 (\bar{S})_{l_1 l_2 \cdots l_n} (D_0 S)_{l_2 l_1 \cdots l_n}
\nonumber \\
&~&  +  \frac{i}{2(n-1)!} \sum_{i_{1}, i_{2}, \cdots, i_{n-1}}
\sum_{l_{1}, l_{2}, \cdots, l_{n}} (\bar{S})_{l_1 l_2 \cdots l_n} 
\gamma^{i_1 i_2 \cdots i_{n-1}} 
[X^{i_1}, \cdots, X^{i_{n-1}}, S]_{l_2 l_1 \cdots l_n} ,
\label{L-susy}
\end{eqnarray}
where $S$ is a Grassmann valued $n$-th power matrix 
and $\gamma^{i_1 i_2 \cdots i_{n-1}}$ is a product of Dirac's $\gamma$ matrices.
This Lagrangian is the counterpart of super $p$-brane 
given by (\ref{S-super-p-brane-lcg-NB}), and the system possesses
supersymmetry between $X^i$ and $S$ for specific numbers of $n$ and $N$.
 (ii) Generalization of matrix model\\
We write down the action of 0-dimensional system analogous to matrix model,
\begin{eqnarray}
&~& S =  \frac{\alpha}{n \cdot n !} \sum_{\mu_{1}, \mu_{2}, \cdots, \mu_{n}}
\sum_{l_{1}, l_{2}, \cdots, l_{n}}
 [X^{\mu_1}, X^{\mu_2}, \cdots, X^{\mu_n}]_{l_1 l_2 \cdots l_n} 
[X^{\mu_1}, X^{\mu_2}, \cdots, X^{\mu_n}]_{l_2 l_1 \cdots l_n} 
\nonumber \\
&~&  - \beta \sum_{\mu}
\sum_{l_{1}, l_{2}, \cdots, l_{n}}
(X^\mu)_{l_1 l_2 \cdots l_n} (X^\mu)_{l_2 l_1 \cdots l_n}
\nonumber \\
&~&  - \gamma \frac{2i}{n+1} \sum_{\mu, \mu_{1}, \mu_{2}, \cdots, \mu_{n}}
\sum_{l_{1}, l_{2}, \cdots, l_{n}}
f^{\mu \mu_1 \mu_2 \cdots \mu_{n}} 
(X^\mu)_{l_1 l_2 \cdots l_n} (X^{\mu_1} X^{\mu_2} \cdots X^{\mu_n})_{l_2 l_1 \cdots l_n} ,
\label{L-model}
\end{eqnarray}
where $X^\mu$s are hermitian $n$-th power matrices and  $\alpha$, $\beta$ and $\gamma$ are
real parameters.
This action with $\beta = \gamma = 0$ is interpreted as 
an $n$-th power matrix analog of the action (\ref{S-p-brane-NB-e}). 
In the case with $n=2$, $\alpha = 1/g^2$ and $\beta = \gamma = 0$,
the action is equivalent to the bosonic part of type IIB matrix model.\cite{IIB}

\section{Conclusions}

We have proposed a generalization of spin algebra using multi-index objects
called $n$-th power matrices and studied
a dynamical system analogous to matrix theory.
We have found that the system has a solution described by generalized spin representation matrices
and possesses a symmetry similar to the volume preserving diffeomorphism 
in the classical $p$-brane action.

Our system is interpreted as a generalization of the bosonic part of the BFSS matrix theory.
The BFSS matrix theory has several interesting physical implications.
For example, the theory is regarded as a regularized theory of supermembrane
or it describes the system of D0-branes and
can offer a microscopic description of M-theory.
The theory takes a special position from the viewpoint of symmetry properties, too.
Our system for $n=2$ has a larger symmetry, that is, the invariance 
under an arbitrary time-dependent unitary transformation, but
it seems to possess a restricted type of local symmetry for $n \geq 3$.
We have treated the abelian local transformations (\ref{G-transf-X}) and (\ref{G-transf-A})
as an example.
We also have considered the case that transformations form a group whose elements are factorized
into a product of matrices, as an extension of unitary transformation.
It is an important subject to explore physical implications
and transformation properties beyond the group theoretical analysis, in our system for $n \geq 3$.

 \section*{Acknowledgements}
This work was supported in part by Scientific Grants from the Ministry of Education
and Science, Nos. 13135217 and 15340078.

\appendix
\section{Definition of $n$-th power matrices}

In this appendix, we define $n$-index objects,
which we refer to as $\lq$$n$-th power matrices',\footnote{
Many-index objects have been introduced 
to construct a quantum version of the Nambu bracket.\cite{many-index,Xiong}
The definition of the $n$-fold product we use is the same as that used by Xiong.}
and define related terminology.\cite{YK-GHD}
An $n$-th power matrix is an object with $n$ indices written $B_{l_1 l_2 \cdots l_n}$.
This is a generalization
of a usual matrix, written analogously as $B_{l_1 l_2}$.
We treat $n$-th power $\lq\lq$square" matrices, i.e., 
$N \times N \times \cdots \times N$ matrices,
and treat the elements of these matrices as $c$-numbers, in many cases,
throughout this paper.

First, we define the hermiticity of an $n$-th power matrix by the relation
$B_{l'_1 l'_2 \cdots l'_n} = B_{l_1 l_2 \cdots l_n}^{*}$ for odd permutations among indices
and refer to an $n$-th power matrix possessing the property of
hermiticity as a $\lq$hermitian $n$-th power matrix'.
Here, the asterisk indicates complex conjugation.
A hermitian $n$-th power matrix satisfies the relation 
$B_{l'_1 l'_2 \cdots l'_n} = B_{l_1 l_2 \cdots l_n}$ for even permutations among indices.
The components for which at least two indices are identical, 
e.g., $B_{l_1 \cdots l_i \cdots l_i \cdots l_n}$,
which is the counterpart of the diagonal part of a hermitian matrix, are real-valued and symmetric
with respect to permutations among indices $\{l_1, \cdots, l_i, \cdots, l_i, \cdots, l_n\}$.
We refer to a special type of hermitian matrix whose components with all different
indices are vanishing 
as a $\lq$real normal form' or a $\lq$real normal $n$-th power matrix'.
A normal $n$-th power matrix is written 
\begin{eqnarray}
B^{(N)}_{l_1 l_2 \cdots l_n} = \sum_{i < j} \delta_{l_i l_j} 
b_{l_j l_1 \cdots \hat{l}_i \cdots \hat{l}_j \cdots l_n} ,
\label{normal} 
\end{eqnarray}
where the summation is over all pairs among $\{l_1, \cdots, l_n\}$, the hatted indices are omitted,
and $b_{l_j l_1 \cdots \hat{l}_i \cdots \hat{l}_j \cdots l_n}$
is symmetric under the exchange of $(n-2)$ indices 
except of $l_j$.

We define the $n$-fold product of $n$-th power matrices $(B_i)_{l_1 l_2 \cdots l_n}$, 
$(i=1,2,...,n)$ by
\begin{eqnarray}
(B_1 B_2 \cdots B_n)_{l_1 l_2 \cdots l_n} \equiv
\sum_k (B_1)_{l_1 \cdots l_{n-1} k} (B_2)_{l_1 \cdots l_{n-2} k l_n} \cdots (B_n)_{k l_2 \cdots l_n} .
\label{product}
\end{eqnarray}
The resultant $n$-index object, $(B_1 B_2 \cdots B_n)_{l_1 l_2 \cdots l_n}$,
is not necessarily hermitian, even if the $n$-th power matrices $(B_i)_{l_1 l_2 \cdots l_n}$
are hermitian.
Note that the above product is, in general, neither commutative nor associative; for example,
we have
\begin{eqnarray}
&~& (B_1 B_2 \cdots B_n)_{l_1 l_2 \cdots l_n} \neq (B_2 B_1 \cdots B_n)_{l_1 l_2 \cdots l_n} , 
\nonumber \\
&~& (B_1 \cdots B_{n-1} (B_n B_{n+1} \cdots B_{2n-1}))_{l_1 l_2 \cdots l_n} 
\nonumber \\
&~& ~~~~~ \neq ((B_1 \cdots B_{n-1} B_n) B_{n+1} \cdots B_{2n-1})_{l_1 l_2 \cdots l_n} .
\end{eqnarray}
The $n$-fold commutator is defined by
\begin{eqnarray}
&~& [B_1, B_2, \cdots, B_n]_{l_1 l_2 \cdots l_n} 
\nonumber \\
&~& ~~~~~~ \equiv 
\sum_{(i_1, \cdots, i_n)} \sum_k \mbox{sgn}(P) 
(B_{i_1})_{l_1 \cdots l_{n-1} k} (B_{i_2})_{l_1 \cdots l_{n-2} k l_n} 
 \cdots (B_{i_n})_{k l_2 \cdots l_n} ,
\label{commutator}
\end{eqnarray}
where the first summation is over all permutations among the subscripts $\{i_1, \cdots, i_n\}$.
Here, sgn($P$) is $+1$ and $-1$ for even and odd permutations 
among the subscripts $\{i_1, \cdots, i_n\}$, respectively.
If the $n$-th power matrices $(B_i)_{l_1 l_2 \cdots l_n}$ are hermitian, then $i[B_1, B_2, \cdots, B_n]_{l_1 l_2 \cdots l_n}$ is also hermitian . 

We study some properties of the $n$-fold commutator 
$[B_1, B_2, \cdots, B_n]_{l_1 l_2 \cdots l_n}$.
This commutator is written
\begin{eqnarray}
[B_1, B_2, \cdots, B_n]_{l_1 l_2 \cdots l_n} 
&=& (B_1)_{l_1 l_2 \cdots l_n} \widetilde{(B_2 B_3 \cdots B_n)}_{l_1 l_2 \cdots l_n} 
\nonumber \\
&~& + (-1)^{n-1} (B_2)_{l_1 l_2 \cdots l_n} \widetilde{(B_3 \cdots B_n B_1)}_{l_1 l_2 \cdots l_n}
\nonumber \\
&~& + \cdots 
 + (-1)^{n-1} (B_n)_{l_1 l_2 \cdots l_n} \widetilde{(B_1 B_2 \cdots B_{n-1})}_{l_1 l_2 \cdots l_n}
\nonumber \\
&~& + ([B_1, B_2, \cdots, B_n])^0_{l_1 l_2 \cdots l_n} ,
\label{commutator2}
\end{eqnarray}
where $\widetilde{(B_2 B_3 \cdots B_{n})}_{l_1 l_2 \cdots l_n}$ 
and $([B_1, B_2, \cdots, B_n])^0_{l_1 l_2 \cdots l_n}$ are defined by
\begin{eqnarray}
&~& \widetilde{(B_2 B_3 \cdots B_{n})}_{l_1 l_2 \cdots l_n} 
\nonumber \\
&~& ~~~ \equiv \sum_{(i_2, \cdots, i_n)} \mbox{sgn}(P) \Bigl( (B_{i_2})_{l_1 \cdots l_{n-2} l_n l_n}
(B_{i_3})_{l_1 \cdots l_{n-3} l_n l_{n-1} l_n} \cdots (B_{i_n})_{l_n l_2 \cdots l_{n-1} l_n}
\nonumber \\
&~& ~~~~ + (-1)^{n-1} (B_{i_2})_{l_1 \cdots l_{n-3} l_{n-1} l_{n-1} l_n}
(B_{i_3})_{l_1 \cdots l_{n-4} l_{n-1} l_{n-2} l_{n-1} l_n} 
\cdots (B_{i_n})_{l_1 \cdots l_{n-2} l_{n-1} l_{n-1}}
\nonumber \\
&~& ~~~~ + \cdots + (-1)^{n-1} (B_{i_2})_{l_1 \cdots l_{n-1} l_1}
(B_{i_3})_{l_1 \cdots l_{n-2} l_{1} l_n} \cdots (B_{i_n})_{l_1 l_1 l_3 \cdots l_n} \Bigr)
\label{widetilde} 
\end{eqnarray}
and 
\begin{eqnarray}
&~& ([B_1, B_2, \cdots, B_n])^0_{l_1 l_2 \cdots l_n} 
\nonumber \\
&~& ~~~~ \equiv 
\sum_{(i_1, \cdots i_n)} \sum_{k \neq l_1, l_2, \cdots, l_n} \mbox{sgn}(P) 
(B_{i_1})_{l_1 \cdots l_{n-1} k} (B_{i_2})_{l_1 \cdots l_{n-2} k l_n} \cdots (B_{i_n})_{k l_2 \cdots l_n} ,
\label{commutator0}
\end{eqnarray}
respectively.

We now discuss features of $\widetilde{(B_1 B_2 \cdots B_{n-1})}_{l_1 l_2 \cdots l_n}$. 
It is skew-symmetric
with respect to permutations among indices, i.e.,
\begin{eqnarray}
&~&  \widetilde{(B_1 B_2 \cdots B_{n-1})}_{l_1 \cdots l_i \cdots l_j \cdots l_n} = 
- \widetilde{(B_1 B_2 \cdots B_{n-1})}_{l_1 \cdots l_j \cdots l_i \cdots l_n} ,
\label{skew-symmetry}
\end{eqnarray}
if each $(B_k)_{l_j l_1 \cdots \hat{l}_i \cdots l_n}$, $(k = 1, \cdots , n-1)$ 
is symmetric with respect to permutations among the $n$-indices 
$\{l_j, l_1, \cdots, \hat{l}_i, \cdots, l_n\}$, as are hermitian $n$-th power matrices.
Here we define the following operation on an $n$-th antisymmetric object $\omega_{m_1 m_2 \cdots m_n}$:
\begin{eqnarray}
(\delta{\omega})_{m_0 m_1 \cdots m_n} \equiv 
\sum_{i=0}^{n} (-1)^{i} {\omega}_{m_0 m_1 \cdots \hat{m}_i \cdots m_n}  ,
\label{delta}
\end{eqnarray}
where the operator $\delta$ is regarded as a coboundary operator 
that changes $n$-th antisymmetric objects into $(n+1)$-th objects,
and it is nilpotent, i.e. $\delta^2(*) =0$.\footnote{
See Ref.~\citen{cohom} for treatments of cohomology.}
If $\omega_{m_1 m_2 \cdots m_k}$ satisfies
the cocycle condition $(\delta \omega)_{m_0 m_1 \cdots m_n} = 0$,
it is called a cocycle.

For arbitrary normal $n$-th power matrices $B_j^{(N)}$,
the $n$-fold commutator among $B$ and $B_j^{(N)}$ is given by 
\begin{eqnarray}
[B_1^{(N)}, \cdots, B_{n-1}^{(N)}, B]_{l_1 l_2 \cdots l_n} 
= (-1)^{n-1} \widetilde{(B_1^{(N)} \cdots B_{n-1}^{(N)})}_{l_1 l_2 \cdots l_n}  B_{l_1 l_2 \cdots l_n} .
\label{formula1}
\end{eqnarray}
If $\widetilde{(B_1^{(N)} \cdots B_{n-1}^{(N)})}_{l_1 l_2 \cdots l_n}$ is a cocycle
for normal $n$-th power matrices $B_j^{(N)}$,
the following fundamental identity holds:
\begin{eqnarray}
&~& [[C_1, \cdots, C_n], B_1^{(N)}, \cdots, B_{n-1}^{(N)}]_{l_1 l_2 \cdots l_n} 
\nonumber\\
&~& ~~~ =  \sum_{i=1}^{n} [C_1, \cdots, [C_i, B_1^{(N)}, 
\cdots, B_{n-1}^{(N)}], \cdots, C_n]_{l_1 l_2 \cdots l_n} .
\label{fundamental}
\end{eqnarray}

Next we give two kinds of trace operation on $n$-th power matrices.
The first one is a generalization of the trace of $B_{l_1 l_2}$ which is defined by
\begin{eqnarray}
\mbox{Tr} B \equiv \sum_{l} B_{ll} = \sum_{l_1, l_2} B_{l_1 l_2} \delta_{l_1 l_2} .
\label{TrB}
\end{eqnarray}
We define the trace operation on the $n$-th power matrix
$B_{l_1 l_2 \cdots l_{n}}$ by
\begin{eqnarray}
\mbox{Tr}_{(1)} B \equiv \sum_{l} B_{l l \cdots l} =
\sum_{l_1, l_2, \cdots, l_{n}} B_{l_1 l_2 \cdots l_{n}} \delta_{l_1 l_2 \cdots l_{n}} ,
\label{Tr(1)B}
\end{eqnarray}
where $\delta_{l_1 l_2 \cdots l_{n}} \equiv \delta_{l_1 l_2} \delta_{l_2 l_3} \cdots \delta_{l_{n-1}l_{n}}$.
The second one is a generalization of the trace of $(B_1 B_2)_{l_1 l_2}$ which is written
\begin{eqnarray}
&~& \mbox{Tr} (B_1 B_2) \equiv \sum_{l_1} \sum_{k} (B_1)_{l_1 k} (B_2)_{k l_1} 
= \sum_{l_1, l_2} (B_1 B_2)_{l_1 l_2} \delta_{l_1 l_2} .
\label{TrB1B2}
\end{eqnarray}
Here we define a new product of $n$-th power matrices $B_1$ and $B_2$ by
\begin{eqnarray}
(B_1 B_2)_{l_1 l_2 \cdots l_{n}} \equiv 
\sum_{k} (B_1)_{l_1 \cdots l_{n-1} k} (B_2)_{l_1 \cdots l_{n-2} k l_{n}} .
\label{B1B2}
\end{eqnarray}
This product is also obtained by setting $B_3 = \cdots = B_n = T$ in the $n$-fold product (\ref{product})
where $T$ is the $n$-th power matrix in which every component 
has the value 1, i.e., $T_{l_1 l_2 \cdots l_n}=1$.
Note that this product is non-commutative but associative;
\begin{eqnarray}
 (B_1 B_2)_{l_1 l_2 \cdots l_n} \neq (B_2 B_1)_{l_1 l_2 \cdots l_n} , 
~~ (B_1 (B_2 B_3))_{l_1 l_2 \cdots l_n} = ((B_1 B_2) B_3)_{l_1 l_2 \cdots l_n} .
\end{eqnarray}
Now we define a trace operation on the $n$-th power matrix
$(B_1 B_2)_{l_1 l_2 \cdots l_{n}}$ by
\begin{eqnarray}
&~& \mbox{Tr}_{(2)} (B_1 B_2) 
\equiv \sum_{l_1, l_2 \cdots, l_{n}} (B_1 B_2)_{l_1 l_2 \cdots l_{n}}
\delta_{l_{n-1} l_{n}} \nonumber \\
&~& ~~~~~~~~~~~~~~~~ = \sum_{l_1, l_2 \cdots, l_{n}} 
\sum_{k} (B_1)_{l_1 \cdots l_{n-1} k} (B_2)_{l_1 \cdots l_{n-2} k l_{n}} \delta_{l_{n-1} l_{n}}
\nonumber \\
&~& ~~~~~~~~~~~~~~~~ = \sum_{l_1, \cdots, l_{n-1}, l_{n}} 
 (B_1)_{l_1 \cdots l_{n-1} l_n} (B_2)_{l_1 \cdots l_{n} l_{n-1}} .
\label{Tr(2)B1B2}
\end{eqnarray}
For a hermitian $n$-th power matrix $B_{l_1 \cdots l_n}$, the second kind of trace 
for $(B^2)_{l_1 l_2 \cdots l_{n}}$
is positive semi-definite;
\begin{eqnarray}
&~& \mbox{Tr}_{(2)} B^2 = \sum_{l_1, \cdots, l_{n}} |B_{l_1 \cdots l_n}|^2 \geq 0 .
\label{Tr(2)BB}
\end{eqnarray}
For hermitian $n$-th power matrices $B_1$ and $B_2$, 
the following formula holds on:
\begin{eqnarray} &~& \mbox{Tr}_{(2)} (B_1 B_2)  = \sum_{l_1, \cdots, l_{n-1}, l_{n}} 
 (B_1)_{l_1 \cdots l_{n-1} l_n} (B_2)_{l_1 \cdots l_{n} l_{n-1}}
\nonumber \\
&~& ~~~~~~~~~~~~~~~~ = \sum_{l_1, l_2 \cdots, l_{n}} 
 (B_1)_{l_1 l_2 \cdots l_n} (B_2)_{l_2 l_1 \cdots l_{n}} .
\label{Tr(2)B1B2-2}
\end{eqnarray}

\section{Transformation properties of $n$-th power matrices}

In this appendix, we study an analog of unitary transformation for $n$-th power matrices.
First we review a unitary transformation for usual matrices.
The unitary transformation for $B_{l_1 l_2}$ is defined by
\begin{eqnarray}
&~& B_{l_1 l_2} \to B'_{l_1 l_2} = \sum_{m_1, m_2} U_{l_1 m_1} B_{m_1 m_2} U_{m_2 l_2}^{\dagger}
 = \sum_{m_1, m_2} U_{l_1 m_1} U_{l_2 m_2}^{*} B_{m_1 m_2} 
\nonumber \\
&~& ~~~~~~~~~~~~~~~~~ \equiv \sum_{m_1, m_2} R_{l_1 l_2}^{m_1 m_2} B_{m_1 m_2} ,
\label{U-transf}
\end{eqnarray}
where $U_{l m}$ is a unitary matrix  ($\sum_{m} U_{lm}U_{mn}^{\dagger} = \sum_m U_{lm}^{\dagger}U_{mn} = \delta_{ln}$)
and $R_{l_1 l_2}^{m_1 m_2}$ is a $\lq\lq$transformation
matrix" defined by $R_{l_1 l_2}^{m_1 m_2}  \equiv U_{l_1 m_1} U_{l_2 m_2}^{*}$.
By definition of $R_{l_1 l_2}^{m_1 m_2}$, we have the relation:
\begin{eqnarray}
&~& (R_{l_1 l_2}^{m_1 m_2})^* = R_{l_2 l_1}^{m_2 m_1} .
\label{R*}
\end{eqnarray}
{}From the unitarity of $U_{lm}$, we obtain the relations:
\begin{eqnarray}
&~& \sum_m R_{l_1 l_2}^{m m} = \delta_{l_1 l_2} , ~~~~~ \sum_l R_{l l}^{m_1 m_2} = \delta_{m_1 m_2} ,
\label{R=d}\\
&~& \sum_k R_{l_1 k}^{m_1 m_2} R_{k l_2}^{n_2 n_1} = R_{l_1 l_2}^{m_1 n_1} \delta_{m_2 n_2} ,
\label{RR=R}\\
&~& \sum_{k,l} R_{l k}^{m_1 m_2} R_{k l}^{n_2 n_1} = \delta_{m_1 n_1} \delta_{m_2 n_2} .
\label{RR=d}
\end{eqnarray}
In terms of $R_{l_1 l_2}^{m_1 m_2}$,
the quantities $\delta_{l_1 l_2}$ and ${\mbox{Tr}}B \equiv \sum_{l} B_{ll}$ are shown 
to be invariant under the unitary transformation from the first relation
and the second one in (\ref{R=d}), respectively.
The relation (\ref{RR=R}) is related to the covariance of 
$C_{l_1 l_2} = (B_1 B_2)_{l_1 l_2}$ under the unitary
transformation,
\begin{eqnarray}
&~& C_{l_1 l_2} \to C'_{l_1 l_2} \equiv (B'_1 B'_2)_{l_1 l_2} =
 \sum_{m_1, m_2} R_{l_1 l_2}^{m_1 m_2} C_{m_1 m_2} .
\label{U-transfC}
\end{eqnarray}
The relation (\ref{RR=d}) is related to the invariance of $(B_1 B_2)_{ll}$ 
under the unitary transformation.

The infinitesimal unitary transformation is given by
\begin{eqnarray}
&~& \delta B_{l_1 l_2} = i[\Lambda, B]_{l_1 l_2}  
= i \sum_{m_1, m_2} {r^{(\Lambda)}}_{l_1 l_2}^{m_1 m_2} B_{m_1 m_2} ,
\label{inf-U(t)-transf-B}
\end{eqnarray}
where the $\lq\lq$transformation matrix" ${r^{(\Lambda)}}_{l_1 l_2}^{m_1 m_2}$
is given by ${r^{(\Lambda)}}_{l_1 l_2}^{m_1 m_2} = \Lambda_{l_1 m_1} \delta_{l_2 m_2} 
 - \Lambda_{m_2 l_2} \delta_{l_1 m_1}$.
We find that the following identity holds among transformation matrices,
\begin{eqnarray}
\sum_{n_1, n_2} ({r^{(\Lambda)}}_{l_1 l_2}^{n_1 n_2} {r^{(\Lambda')}}_{n_1 n_2}^{m_1 m_2}
 - {r^{(\Lambda')}}_{l_1 l_2}^{n_1 n_2} {r^{(\Lambda)}}_{n_1 n_2}^{m_1 m_2})
 = {r^{([\Lambda, \Lambda'])}}_{l_1 l_2}^{m_1 m_2} 
\label{r-identity}
\end{eqnarray}
{}from the Jacobi identity.
We can also show that the commutator $[B_1, B_2]$ transforms as
\begin{eqnarray}
&~& \delta [B_1, B_2]_{l_1 l_2} = [\delta B_1, B_2]_{l_1 l_2}
+ [B_1, \delta B_2]_{l_1 l_2}
\nonumber \\
&~& ~~~~~~ = i[\Lambda, [B_1, B_2]]_{l_1 l_2} 
= i \sum_{m_1, m_2} {r^{(\Lambda)}}_{l_1 l_2}^{m_1 m_2} B_{m_1 m_2} .
\label{inf-U(t)-transf-[B1,B2]}
\end{eqnarray}

Next we study the case of $n$-th power matrices with $n \geq 3$.
We define an extension of the unitary transformation (\ref{U-transf}) for $B_{l_1 l_2 \cdots l_n}$ by
\begin{eqnarray}
&~&  B_{l_1 l_2 \cdots l_n} \to B'_{l_1 l_2 \cdots l_n} 
= \sum_{m_1, m_2, \cdots, m_n} R_{l_1 l_2 \cdots l_n}^{m_1 m_2 \cdots m_n} B_{m_1 m_2 \cdots m_n} ,
\label{R-transf}
\end{eqnarray}
where $R_{l_1 l_2 \cdots l_n}^{m_1 m_2 \cdots m_n}$ is a $\lq\lq$transformation matrix". 
{}From the transformation (\ref{R-transf}) and the hermiticity of $B_{l_1 l_2 \cdots l_n}$,
we obtain the relations:
\begin{eqnarray}
(R_{l'_1 l'_2 \cdots l'_n}^{m'_1 m'_2 \cdots m'_n})^* = R_{l_1 l_2 \cdots l_n}^{m_1 m_2 \cdots m_n}
\label{R-permute-odd}
\end{eqnarray}
for odd permutations among the pairs of indices $(l_k, m_k)$, $(k=1, \cdots, n)$
and 
\begin{eqnarray}
(R_{l'_1 l'_2 \cdots l'_n}^{m'_1 m'_2 \cdots m'_n}) = R_{l_1 l_2 \cdots l_n}^{m_1 m_2 \cdots m_n}
\label{R-permute-even}
\end{eqnarray}
for even permutations among the pairs of indices $(l_k, m_k)$, $(k=1, \cdots, n)$.
When the transformation (\ref{R-transf}) is given by the $n$-fold product,
transformations do not necessarily form a group because the $n$-fold product
is, in general, non-associative.
For simplicity, here we treat the case that transformations form a group 
and the transformation matrix $R_{l_1 l_2 \cdots l_n}^{m_1 m_2 \cdots m_n}$
is factorized into a product of matrices such that
\begin{eqnarray}
R_{l_1 l_2 \cdots l_n}^{m_1 m_2 \cdots m_n} 
= V_{l_1}^{m_1} V_{l_2}^{m_2} \cdots V_{l_n}^{m_n} ,
\label{RVn}
\end{eqnarray}
where $V_l^m$ should be a real matrix from the relations (\ref{R-permute-odd}) and (\ref{R-permute-even}).
The form of matrix $V_l^m$ is restricted by suitable requirements.
Let us give examples.
The first requirement is that the second kind of trace 
$\mbox{Tr}_{(2)} (BC) = \sum_{l_1,l_2, \cdots, l_n} B_{l_1 l_2 \cdots l_n} C_{l_2 l_1 \cdots l_n}$ is
invariant under the transformation:
\begin{eqnarray}
&~& B_{l_1 l_2 \cdots l_n} \to B'_{l_1 l_2 \cdots l_n} 
= \sum_{m_1, m_2, \cdots, m_n} R_{l_1 l_2 \cdots l_n}^{m_1 m_2 \cdots m_n} B_{m_1 m_2 \cdots m_n} , 
\nonumber \\
&~& C_{l_1 l_2 \cdots l_n} \to C'_{l_1 l_2 \cdots l_n} 
= \sum_{m_1, m_2, \cdots, m_n} R_{l_1 l_2 \cdots l_n}^{m_1 m_2 \cdots m_n} C_{m_1 m_2 \cdots m_n} .
\label{R-transfBCn}
\end{eqnarray}
The necessary condition is given by
\begin{eqnarray}
\sum_{l_1, l_2, \cdots, l_n} 
R_{l_1 l_2 \cdots l_n}^{m_1 m_2 \cdots m_n} R_{l_2 l_1 \cdots l_n}^{m'_2 m'_1 \cdots m'_n} 
= \delta_{m_1 m'_1} \delta_{m_2 m'_2} \cdots \delta_{m_n m'_n} 
\label{RR=dn}
\end{eqnarray} 
and then $R_{l_1 l_2 \cdots l_n}^{m_1 m_2 \cdots m_n}$ is given by
\begin{eqnarray}
R_{l_1 l_2 \cdots l_n}^{m_1 m_2 \cdots m_n} 
= O_{l_1}^{m_1} O_{l_2}^{m_2} \cdots O_{l_n}^{m_n} ,
\label{ROn}
\end{eqnarray}
where $O_l^m$ is an orthogonal matrix
($\sum_l O_l^m O_l^n = \delta_{mn}$, $\sum_m O_l^m O_n^m = \delta_{ln}$). 

Next we require that the trace
$\mbox{Tr}_{(2)} C^2 = \sum_{l_1, l_2, \cdots, l_n} C_{l_1 l_2 \cdots l_n} C_{l_2 l_1 \cdots l_n}$ is
invariant under the transformation:
\begin{eqnarray}
&~& (B_k)_{l_1 l_2 \cdots l_n} \to (B_k)'_{l_1 l_2 \cdots l_n} 
= \sum_{m_1, m_2, \cdots, m_n} R_{l_1 l_2 \cdots l_n}^{m_1 m_2 \cdots m_n} (B_k)_{m_1 m_2 \cdots m_n} ,
\label{R-transfBk}
\end{eqnarray}
where $C_{l_1 l_2 \cdots l_n}$ is the $n$-fold product given by
$C_{l_1 l_2 \cdots l_n}=(B_1 B_2 \cdots B_n)_{l_1 l_2 \cdots l_n}$.
We find that the following $R_{l_1 l_2 \cdots l_n}^{m_1 m_2 \cdots m_n}$ satisfies 
the above requirement,
\begin{eqnarray}
R_{l_1 l_2 \cdots l_n}^{m_1 m_2 \cdots m_n} =
 D_{l_1}^{m_1} D_{l_2}^{m_2} \cdots D_{l_n}^{m_n}  ,
\label{RDn}
\end{eqnarray}
where $D_{l}^{m} = \delta_{l}^{\sigma(m)}$
and $\sigma(m)$ stands for a permutation of the index $m$.
Note that $D_l^i$ satisfies the relations:
\begin{eqnarray}
&~& \sum_i D_{m_1}^i D_{m_2}^i \cdots D_{m_k}^i = \delta_{m_1 m_2 \cdots m_k} ,
~~~ \sum_m D_m^{i_1} D_m^{i_2} \cdots D_m^{i_k} = \delta_{i_1 i_2 \cdots i_k} 
\label{D-rel}
\end{eqnarray}
for an arbitrary integer $k$.
Using the relations (\ref{D-rel}) with $k=n$, we find that 
$\delta_{l_1 l_2 \cdots l_n}$ and $\mbox{Tr}_{(1)}B$ are invariant
under the extended transformation (\ref{R-transf}). 
Further we find that the $n$-fold product $C_{l_1 l_2 \cdots l_n}=(B_1 B_2 \cdots B_n)_{l_1 l_2 \cdots l_n}$ 
transforms covariantly,
\begin{eqnarray}
&~&  C_{l_1 l_2 \cdots l_n} \to C'_{l_1 l_2 \cdots l_n} 
= \sum_{m_1, m_2, \cdots, m_n} R_{l_1 l_2 \cdots l_n}^{m_1 m_2 \cdots m_n} C_{m_1 m_2 \cdots m_n} 
\label{R-transfB}
\end{eqnarray}
under the transformation (\ref{R-transfBk}) with the transformation matrix (\ref{RDn}).

Finally we study a generalization of infinitesimal transformation (\ref{inf-U(t)-transf-B})
for $n$-th power matrices with $n \geq 3$.
We consider the following transformation
by the use of $n$-fold commutator,
\begin{eqnarray}
&~& \delta B_{l_1 l_2 \cdots l_n} 
= i[\Lambda_1, \cdots, \Lambda_{n-1}, B]_{l_1 l_2 \cdots l_n}  
\nonumber \\ 
&~& ~~~~~~~~~ = i \sum_{m_1, m_2, \cdots, m_n} 
 {r^{(\Lambda)}}_{l_1 l_2 \cdots l_n}^{m_1 m_2 \cdots m_n} 
 B_{m_1 m_2 \cdots m_n} ,
\label{inf-U(t)-transf-B-n}
\end{eqnarray}
where ${r^{(\Lambda)}}_{l_1 l_2 \cdots l_n}^{m_1 m_2 \cdots m_n}$ is 
a $\lq\lq$transformation matrix" and $\Lambda_k$, $(k=1, \cdots, n)$ is a set of $\lq\lq$generators".
The $n$-fold commutator $[B_1, B_2, \cdots, B_n]$ transforms as
\begin{eqnarray}
&~& \delta [B_1, B_2, \cdots, B_n]_{l_1 l_2 \cdots l_n} 
= \sum_{k=1}^n [B_1, \cdots, \delta(B_k), \cdots, B_n]_{l_1 l_2 \cdots l_n} 
\nonumber \\
&~& ~~ = i \sum_p \sum_{(i_1, \cdots, i_n)} \sum_k \mbox{sgn}(P) 
(B_{i_1})_{l_1 \cdots l_{n-1} k}  
\nonumber \\
&~& ~~~~ \cdots \sum_{m_1, m_2, \cdots, m_n} 
{r^{(\Lambda)}}_{l_1 \cdots l_{n-p} k l_{n+2-p} \cdots l_n}^{m_1 m_2 \cdots m_n} 
(B_{i_p})_{m_1 m_2 \cdots m_n} 
 \cdots (B_{i_n})_{k l_2 \cdots l_n} 
\label{inf-U(t)-transf-Bcommutator-n}
\end{eqnarray}
under the transformation:
\begin{eqnarray}
&~& \delta (B_k)_{l_1 l_2 \cdots l_n} = i \sum_{m_1, m_2, \cdots, m_n} 
 {r^{(\Lambda)}}_{l_1 l_2 \cdots l_n}^{m_1 m_2 \cdots m_n} 
 (B_k)_{m_1 m_2 \cdots m_n} .
\label{inf-U(t)-transf-Bk-n}
\end{eqnarray}
Note that the following transformation law does not 
necessarily hold 
for $n$-th power matrices with $n \geq 3$,
\begin{eqnarray} &~& \delta [B_1, B_2, \cdots, B_n]_{l_1 l_2 \cdots l_n} 
\nonumber \\
&~& ~~~~~~~~~~~~~ = i \sum_{m_1, m_2, \cdots, m_n} 
{r^{(\Lambda)}}_{l_1 l_2 \cdots l_n}^{m_1 m_2 \cdots m_n} 
 [B_1, B_2, \cdots, B_n]_{m_1 m_2 \cdots m_n} .
\label{inf-U(t)-transf-commB-n}
\end{eqnarray}
This fact means that the fundamental identity does not always hold among $n$-th power matrices
with $n \geq 3$.
Here we treat the case that $\Lambda_k$, $(k=1, 2, \cdots, n)$ are normal $n$-th power matrices
as an example that the above law (\ref{inf-U(t)-transf-commB-n}) holds.
In this case, the $n$-th power matrices $B_k$ transform as
\begin{eqnarray}
&~& \delta (B_k)_{l_1 l_2 \cdots l_n} 
= i[\Lambda_1, \cdots, \Lambda_{n-1}, B_k]_{l_1 l_2 \cdots l_n}  
= i \Lambda_{l_1 l_2 \cdots l_n} (B_k)_{l_1 l_2 \cdots l_n} ,
\label{inf-U(t)-transf-B-normal}
\end{eqnarray}
where $\Lambda_{l_1 l_2 \cdots l_n} 
= (-1)^{n-1}(\widetilde{\Lambda_1 \cdots \Lambda_{n-1}})_{l_1 l_2 \cdots l_n}$.
In terms of $\Lambda_{l_1 l_2 \cdots l_n}$, the transformation matrix 
${r^{(\Lambda)}}$ is written 
\begin{eqnarray}
{r^{(\Lambda)}}_{l_1 l_2 \cdots l_n}^{m_1 m_2 \cdots m_n}
= \Lambda_{m_1 m_2 \cdots m_n} \delta_{l_1}^{m_1} \delta_{l_2}^{m_2} \cdots
\delta_{l_n}^{m_n} .
\label{r-Lambda}
\end{eqnarray}
If $\Lambda_{l_1 l_2 \cdots l_n}$s satisfy the cocycle condition:
\begin{eqnarray}
(\delta \Lambda)_{l_0 l_1 \cdots l_n} \equiv 
\sum_{i=0}^{n} (-1)^{i} \Lambda_{l_0 l_1 \cdots \hat{l}_i \cdots l_n} = 0 ,
\end{eqnarray}
we find that the $n$-fold commutator $[B_1, B_2, \cdots, B_n]$ transforms covariantly:
\begin{eqnarray}
&~& \delta [B_1, B_2, \cdots, B_n]_{l_1 l_2 \cdots l_n} 
= i \Lambda_{l_1 l_2 \cdots l_n} [B_1, B_2, \cdots, B_n]_{l_1 l_2 \cdots l_n} .
\label{inf-U(t)-transf-commB-normal}
\end{eqnarray}

\section{Classical analog of generalized spin algebra}

In this appendix, we study the classical analog of generalized spin algebra.
First we consider the following action integral whose variables are $\phi^i(t)$,\cite{non-bos}
\begin{eqnarray}
S = \int \big(\sum_i A_i(\phi) \frac{d \phi^i}{d t} - H(\phi) \big)dt .
\label{AI-1}
\end{eqnarray}
The change in $S$ under the infinitesimal variation of $\phi^i(t)$ is given by
\begin{eqnarray}
\delta S = \int \sum_{i} 
\big(\sum_{j} F_{ij}(\phi) \frac{d \phi^j}{d t} - \frac{\partial H}{\partial \phi^i}\big)
\delta \phi^i dt ,
\label{deltaAI-1}
\end{eqnarray}
where $F_{ij}(\phi)$ is defined by
\begin{eqnarray}
F_{ij}(\phi) \equiv \frac{\partial A_j}{\partial \phi^i} - \frac{\partial A_i}{\partial \phi^j} .
\label{Fij}
\end{eqnarray}
{}From the least action principle, we obtain the equation of motion:
\begin{eqnarray}
\frac{d \phi^i}{d t} = \sum_{j} F^{ij}\frac{\partial H}{\partial \phi^j}  ,
\label{phi-EOM}
\end{eqnarray}
where $F^{ij}$ is the inverse matrix of $F_{ij}$, i.e., $\sum_{j} F^{ij}F_{jk} = \delta^i_k$.
Then the Poisson bracket is defined by
\begin{eqnarray}
\{f, g\}_{\rm PB} \equiv  \sum_{i, j} F^{ij}\frac{\partial f}{\partial \phi^i} 
\frac{\partial g}{\partial \phi^j} 
\label{PB}
\end{eqnarray}
and, by the use of it, the equation of motion (\ref{phi-EOM}) is rewritten 
\begin{eqnarray}
\frac{d \phi^i}{d t} = \{\phi^i, H\}_{\rm PB} .
\label{phi-EOM2}
\end{eqnarray}
When variables compose of a triplet $X^i$, $(i = 1, 2, 3)$ 
and $F^{ij} = \sum_k \varepsilon^{ijk} X^k$,
$X^i$s satisfy the algebra: \begin{eqnarray}
\{X^i, X^j\}_{\rm PB} = \sum_k \varepsilon^{ijk} X^k .
\label{ClSpin}
\end{eqnarray}
This is the classical analog of the spin algebra ${\it su}(2)$.
In this case, the first term of the action integral (\ref{AI-1})
is rewritten as  \begin{eqnarray}
\int \sum_i A_i(X) d X^i = \frac{1}{2} \int\int \sum_{i,j} F_{ij}(X) d X^i \wedge d X^j 
= \int\int R \sin\theta d\theta \wedge d\varphi ,
\label{AI-1-S2}
\end{eqnarray}
where $\wedge$ represents Cartan's wedge product,
$X^i$s are coordinates on $S^2$ with radius $R$ and they are written
by using the polar coordinates:
\begin{eqnarray}
X^1 = R \sin\theta \cos\varphi , ~~ X^2 = R \sin\theta \sin\varphi , ~~ X^3 = R \cos\theta .
\label{polar}
\end{eqnarray}
The action integral (\ref{AI-1-S2}) is regarded as an area on $S^2$.

Next we consider a generalization of the action integral (\ref{AI-1})
whose variables are $\phi^i=\phi^i(t, \sigma_1, \cdots, \sigma_{n-1})$,
\begin{eqnarray}
&~& S = \int\cdots\int 
\big(\sum_{i_1, \cdots, i_{n-1}, i_n}} {A_{i_1 \cdots i_{n-1} i_n}(\phi) 
\frac{\partial \phi^{i_1}}{\partial \sigma_1}\cdots
\frac{\partial \phi^{i_{n-1}}}{\partial \sigma_{n-1}}\frac{\partial \phi^{i_n}}{\partial t}
\nonumber \\
&~& ~~~~~~~~~~~~~~~~~~ - H_1 \frac{\partial(H_2, \cdots, H_n)}
{\partial(\sigma_1, \cdots, \sigma_{n-1})} \big)
d\sigma_1\cdots d\sigma_{n-1} dt ,
\label{AI-n}
\end{eqnarray}
where $A_{i_1 \cdots i_n}(\phi)$ is antisymmetric under the exchange of indices
and $H_i$s are $\lq\lq$ Hamiltonians".
The change in $S$ under the infinitesimal variation of $\phi^i$ is given by
\begin{eqnarray}
&~& \delta S = \int\cdots\int \sum_{j,i_1, \cdots, i_{n-1}}
\big(\sum_{i_n} F_{ji_1 \cdots i_{n-1} i_n} \frac{\partial \phi^{i_n}}{\partial t} 
\nonumber \\
&~& ~~~~~~~~~~~~~ - \frac{\partial(H_1, H_2, \cdots, H_n)}
{\partial(\phi^j, \phi^{i_1} \cdots, \phi^{i_{n-1}})}\big)
\delta \phi^j  \frac{\partial \phi^{i_1}}{\partial \sigma_1}\cdots
\frac{\partial \phi^{i_{n-1}}}{\partial \sigma_{n-1}} d\sigma_1\cdots d\sigma_{n-1} dt ,
\label{deltaAI-n}
\end{eqnarray}
where $F_{ji_1 \cdots i_n}$ is defined by
\begin{eqnarray}
F_{ji_1 \cdots i_n} \equiv \frac{\partial A_{i_1\cdots i_n}}{\partial \phi^j} 
 + (-1)^n \frac{\partial A_{i_2 \cdots i_n j}}{\partial \phi^{i_1}} 
 + \cdots +  (-1)^n \frac{\partial A_{ji_1 \cdots i_{n-1}}}{\partial \phi^{i_n}}.
\label{Fn}
\end{eqnarray}
{}From the least action principle, we obtain the equation of motion:
\begin{eqnarray}
\frac{d \phi^i}{d t} = \sum_{i_1, \cdots, i_n} F^{ii_1 \cdots i_{n}} 
\frac{\partial(H_1, H_2, \cdots, H_n)}{\partial(\phi^{i_1} \cdots, \phi^{i_{n}})} ,
\label{phi-EOMn}
\end{eqnarray}
where $F^{ii_1 \cdots i_{n}}$ is the inverse of $F_{ii_1 \cdots i_{n}}$,  i.e., $\sum_{i_1, \cdots, i_n} F^{ii_1 \cdots i_{n}}F_{i_1 \cdots i_{n}j} = \delta^i_j$.
Then the Nambu bracket is defined by
\begin{eqnarray}
\{f_1, \cdots, f_{n+1}\}_{\rm NB} \equiv  
\sum_{i_1, \cdots, i_{n+1}} F^{i_1 \cdots i_{n+1}}\frac{\partial f_1}{\partial \phi^{i_1}} \cdots 
\frac{\partial f_{n+1}}{\partial \phi^{i_{n+1}}} 
\label{NB} \end{eqnarray}
and, by the use of it, the equation of motion (\ref{phi-EOMn}) is rewritten 
\begin{eqnarray}
\frac{d \phi^i}{d t} = \{\phi^i, H_1, \cdots, H_n\}_{\rm NB} .
\label{phi-EOM2n}
\end{eqnarray}
The equation of motion (\ref{phi-EOM2n}) is equivalent to the Hamilton-Nambu equation.\cite{Nambu}
When variables compose of an $(n+2)$-let $X^i$, $(i = 1, \cdots, n+2)$ 
and $F^{i_1 \cdots i_{n+1}} = \sum_{i_{n+2}} \varepsilon^{i_1 \cdots i_{n+1} i_{n+2}} X^{i_{n+2}}$,
$X^i$s satisfy the algebra:
\begin{eqnarray}
\{X^{i_1}, \cdots, X^{i_{n+1}}\}_{\rm NB} = \sum_{i_{n+2}}
\varepsilon^{i_1 \cdots i_{n+1} i_{n+2}} X^{i_{n+2}} .
\label{ClgSpin}
\end{eqnarray}
This is the classical analog of the generalized spin algebra.
In this case, the first term of the action integral (\ref{AI-n})
is rewritten as
\begin{eqnarray}
&~& \frac{1}{n!} \int \cdots \int \sum_{i_1, \cdots, i_n}
 A_{i_1 \cdots i_n}(X) dX^{i_1} \wedge \cdots \wedge dX^{i_n} \nonumber \\
&~& = \frac{1}{(n+1)!} \int \cdots \int \sum_{i_1, \cdots, i_{n+1}} F_{i_1 \cdots i_{n+1}}(X) 
dX^{i_1} \wedge \cdots \wedge dX^{i_{n+1}}
\nonumber \\
&~&  = \int\cdots\int R^{n} \sin\theta_2 \sin^2\theta_3 \cdots \sin^n\theta_{n+1}
d\theta_2 \wedge d\theta_1 \wedge d\theta_3 \wedge \cdots \wedge d\theta_{n+1} ,
\label{AI-1-Sn+1}
\end{eqnarray}
where $X^i$s are coordinates on $S^{n+1}$ with radius $R$ and they are written by using the polar coordinates:
\begin{eqnarray}
&~& X^1 = R \sin\theta_{n+1} \sin\theta_{n} \cdots \sin\theta_3 \sin\theta_2 \cos\theta_1 , \\
\nonumber
&~& X^2 = R \sin\theta_{n+1} \sin\theta_{n} \cdots \sin\theta_3 \sin\theta_2 \sin\theta_1, \\
\nonumber
&~& X^3 = R \sin\theta_{n+1} \sin\theta_{n} \cdots \sin\theta_3 \cos\theta_2 ,
~~ \cdots,  X^{n+1} = R \sin\theta_{n+1} \cos\theta_n , \\
\nonumber
&~& X^{n+2} = R \cos\theta_{n+1} .
\label{Sn+1}
\end{eqnarray}
The action integral (\ref{AI-1-Sn+1}) is regarded as an $\lq\lq$area" on $S^{n+1}$.

\section{Classical analog of generalized matrix systems}

In this appendix, we explain the framework of classical $p$-branes.
The bosonic $p$-brane action is given by\cite{p-brane}
\begin{eqnarray}
 S =  - \int d^{p+1}\sigma \sqrt{-g} ,
\label{S-p-brane}
\end{eqnarray}
where $d^{p+1}\sigma$ represents the $(p+1)$-dimensional 
world-volume element and $g = \mbox{det}g_{\alpha \beta}$.
Here $g_{\alpha \beta}$ is the induced world-volume metric given by
\begin{eqnarray}
 g_{\alpha \beta} =\sum_{\mu, \nu} \eta_{\mu \nu} \frac{\partial X^\mu}{\partial \sigma^\alpha} 
 \frac{\partial X^\nu}{\partial \sigma^\beta} ,
\label{g-metric}
\end{eqnarray}
where $X^\mu$, $(\mu = 0, 1, \cdots, D-1)$ are the target space coordinates of $p$-brane and
$\sigma^\alpha$, $(\alpha =0, 1, \cdots, p)$ are the $(p+1)$-dimensional world-volume coordinates.
We assume that the target space is the $D$-dimensional Minkowski space. 
The action integral (\ref{S-p-brane}) is invariant under the reparametrization:
\begin{eqnarray}
 \delta X^\mu = \sum_{\alpha} \epsilon^\alpha \partial_\alpha X^\mu ,
\label{repara}
\end{eqnarray}
where $\epsilon^\alpha$ is an arbitrary function of $\sigma^\alpha$.

Let us introduce the light-cone coordinates in space-time:
\begin{eqnarray}
X^{\pm} = \frac{1}{\sqrt{2}} (X^0 \pm X^{D-1}) .
\label{Xpm}
\end{eqnarray}
The transverse coordinates are denoted by $X^i$, $(i = 1, \cdots, D-2)$.
By using the reparametrization invariance, we can choose the light-cone gauge:
\begin{eqnarray}
X^{+} = x^+ + p^+ \tau ,
\label{lcg}
\end{eqnarray}
where $x^+$ and $p^+$ are the center of mass position and momemtum, respectively,
and $\tau = \sigma^0$.
In the light-cone gauge, the action is written by (up to a zero mode term)
\begin{eqnarray}
 S =  \frac{1}{2} \int d^{p+1}\sigma 
\left(\sum_i (D_0 X^i)^2 - \mbox{det}g_{ab}\right) ,
\label{S-p-brane-lcg}
\end{eqnarray}
where $g_{a b}$ is the induced $p$-dimensional metric given by
\begin{eqnarray}
 g_{a b} = \sum_{i, j} \eta_{i j} \frac{\partial X^i}{\partial \sigma^a} 
 \frac{\partial X^j}{\partial \sigma^b} .
\label{g-metric-p}
\end{eqnarray}
Here $\sigma^a$, $(a =1, \cdots, p)$ are the $p$-dimensional volume coordinates.
The covariant time-derivative $D_0$ is defined by
\begin{eqnarray}
 D_0 X^i \equiv  \left(\frac{\partial}{\partial \tau} 
+ \sum_a u^a \frac{\partial}{\partial \sigma^a}\right) X^i ,
\label{D0X-p}
\end{eqnarray}
where $u^a$ is regarded as the $\lq\lq$gauge field' of time
which should satisfy the equation:
\begin{eqnarray}
 \sum_a \frac{\partial u^a}{\partial \sigma^a} = 0 .
\label{du=0}
\end{eqnarray}
The action (\ref{S-p-brane-lcg}) is rewritten
\begin{eqnarray}
 S =  \frac{1}{2} \int d^{p+1}\sigma 
\left(\sum_i (D_0 X^i)^2 - \frac{1}{p!}\sum_{i_{1}, \cdots, i_{p}} \{X^{i_1}, \cdots, X^{i_p}\}^2\right) ,
\label{S-p-brane-lcg-NB}
\end{eqnarray}
where the symbol $\{f_1, \cdots, f_p\}$ is defined by
\begin{eqnarray}
 \{f_1, \cdots, f_p\} \equiv \sum_{a_1, \cdots, a_{p}} \varepsilon^{a_1 \cdots a_{p}}
 \frac{\partial f_1}{\partial \sigma^{a_1}} \cdots \frac{\partial f_{p}}{\partial \sigma^{a_{p}}} .
\label{symbol}
\end{eqnarray}
(If $\sigma^i$s form a canonical $p$-let, the symbol $\{f_1, \cdots, f_p\}$
is regarded as the Nambu bracket.)
In the case that $u^a$ is written in terms of functions $A_k$, $(k=1, \cdots, p-1)$ as
\begin{eqnarray}
 u^a = \sum_{a_1, \cdots, a_{p-1}} \varepsilon^{a_1 \cdots a_{p-1} a}
 \frac{\partial A_1}{\partial \sigma^{a_1}} \cdots \frac{\partial A_{p-1}}{\partial \sigma^{a_{p-1}}} ,
\label{ua}
\end{eqnarray}
the covariant time-derivative (\ref{D0X-p}) can be written by
\begin{eqnarray}
 D_0 X^i =  \frac{\partial X^i}{\partial \tau} 
+ \{A_1, \cdots, A_{p-1}, X^{i}\} .
\label{D0X-p-2}
\end{eqnarray}
that the action (\ref{S-p-brane-lcg-NB}) is invariant under the $p$-dimensional volume preserving diffeomorphism:
\begin{eqnarray}
&~& \delta X^i = \sum_{a} \lambda^a \partial_a X^i ,
\label{VPD-X}\\
&~& \delta u^a = - \frac{\partial \lambda^a}{\partial \tau} - \sum_b u^b \partial_b  
\lambda^a + \sum_{b} \lambda^b \partial_b u^a ,
\label{VPD-ua}
\end{eqnarray}
where $\lambda^a$ satisfies the condition:
\begin{eqnarray}
 \sum_a \frac{\partial \lambda^a}{\partial \sigma^a} = 0 .
\label{dlambda=0}
\end{eqnarray}
In the case that $\lambda^a$ is written 
in terms of functions $\Lambda_k$, $(k=1, \cdots, p-1)$ as
\begin{eqnarray}
 \lambda^a = \sum_{a_1, \cdots, a_{p-1}} \varepsilon^{a_1 \cdots a_{p-1} a}
 \frac{\partial \Lambda_1}{\partial \sigma^{a_1}} 
\cdots \frac{\partial \Lambda_{p-1}}{\partial \sigma^{a_{p-1}}} ,
\label{lambdaa}
\end{eqnarray}
the transformation laws of $p$-dimensional volume preserving diffeomorphism (\ref{VPD-X}) and (\ref{VPD-ua})
are rewritten as
\begin{eqnarray}
&~& \delta X^i = \{\Lambda_1, \cdots, \Lambda_{p-1}, X^i\} ,
\label{VPD-X-NB}\\
&~& \delta u^a = - \frac{\partial \lambda^a}{\partial \tau} 
 - \{A_1, \cdots, A_{p-1}, \lambda^a\} + \{\Lambda_1, \cdots, \Lambda_{p-1}, u^a\} ,
\label{VPD-ua-NB}
\end{eqnarray}
respectively.
The system has the extra symmetry that $u^a = \sum_b \partial_b W^{ab}$ 
is invariant under the transformation:
\begin{eqnarray}
&~& W^{ab} 
\left(= \sum_{a_1, \cdots, a_{p-2}} \varepsilon^{a_1 \cdots a_{p-2} b a}
 \frac{\partial A_1}{\partial \sigma^{a_1}} \cdots 
 \frac{\partial A_{p-2}}{\partial \sigma^{a_{p-2}}} A_{p-1}\right) \nonumber \\
&~& ~~~~~~ \to {W'}^{ab} = W^{ab} + \sum_c \partial_c \Theta^{abc} ,
\label{W-transf}
\end{eqnarray}
where $\Theta^{abc}$ is an arbitrary antisymmetric function of $\sigma^a$.

Next we write down the action integral of super $p$-brane:
\begin{eqnarray}
&~& S =  \frac{1}{2} \int d^{p+1}\sigma 
\left(\sum_i (D_0 X^i)^2 - \frac{1}{p!}\sum_{i_{1}, \cdots, i_{p}} \{X^{i_1}, \cdots, X^{i_p}\}^2\right.
\nonumber \\
&~& ~~~~ \left. + \frac{i}{2} \bar{S} D_0 S + \frac{1}{(p-1)!}
\sum_{i_{1}, \cdots, i_{p-1}} \bar{S} \gamma^{i_1 \cdots i_{p-1}} \{X^{i_1}, \cdots, X^{i_{p-1}}, S\}^2\right) ,
\label{S-super-p-brane-lcg-NB}
\end{eqnarray}
where $S$ in the right hand side is a spinor of $SO(d-2)$ 
and $\gamma^{i_1 i_2 \cdots i_{n-1}}$ is a product of Dirac's $\gamma$ matrices.
It is well-known that super $p$-branes exist in special dimensions of space-time.\cite{brane-scan}

Finally we discuss an alternative formulation of bosonic $p$-branes.
The action (\ref{S-p-brane}) is rewritten as
\begin{eqnarray}
 S =  - \int d^{p+1}\sigma \sqrt{\frac{1}{(p+1)!}\sum_{\mu_{1}, \cdots, \mu_{p+1}}
\{X^{\mu_1}, \cdots, X^{\mu_{p+1}}\}^2} .
\label{S-p-brane-NB} \end{eqnarray}
By introducing the auxiliary field $e=e(\sigma)$, we can write down the following action 
which is classically equivalent to the above action (\ref{S-p-brane-NB}),
\begin{eqnarray}
 S =  \frac{1}{2} \int d^{p+1}\sigma 
\left(\frac{1}{(p+1)!e}\sum_{\mu_{1}, \cdots, \mu_{p+1}}\{X^{\mu_1}, \cdots, X^{\mu_{p+1}}\}^2 -e\right) .
\label{S-p-brane-NB-e}
\end{eqnarray}
The action (\ref{S-p-brane-NB-e}) is a $p$-brane generalization of so-called Schild action.\cite{Schild}


\begin{thebibliography}{99}

\bibitem{Wigner}
E.P.~Wigner, {\it Group Theory and its Application to the Quantum Mechanics
of Atomic Spectra} (Academic Press, 1959).

\bibitem{Connes}
A.~Connes, {\it Noncommutative geometry} (Academic Press, 1994).

\bibitem{FuzzyS2}
J.~Madore, \PL{B263,1991,245}; Class. and Quant. Grav. \andvol{9,1992,69}.

\bibitem{dWHN}
B.~de~Wit, J.~Hoppe and H.~Nicolai, \NP{B305,1988,545}.

\bibitem{sph-mem}
D.~Kabat and I.W.~Taylor, Adv. Theor. Math. {2,1998,181}.\\
S.J.~Rey, hep-th/9711081.

\bibitem{Myers}
R.C.~Myers, J. High Energy Phys. \andvol{12,1999,022}.

\bibitem{Model-CS} S.~Iso, Y.~Kimura, K.~Tanaka and K.~Wakatsuki, \NP{B604,2001,121}.

\bibitem{GKP}
H.~Grosse, C.~Klimicik and P.~Presnajder, \CMP{180,1996,429}.

\bibitem{FuzzyS2k}
J.~Castelino, S.~Lee and W.~Taylor, \NP{B526,1998,334}.\\ 
S.~Ramgoolam, \NP{B610,2001,461}.\\
P.M.~Ho and S.~Ramgoolam, \NP{B627,2002,266}.

\bibitem{QHE}
S.C.~ Zhang and J.P.~Hu, Science \andvol{294,2001,823}.\\
J.P.~Hu and S.C.~Zhang, cond-mat/0112432.\\
M.~Fabinger, J. High Energy Phys. \andvol{05,2002,037}.\\
K. Hasebe and Y.~Kimura, \PL{B602,2004,255}.

\bibitem{FuzzyCoset}
P.~Aschieri, J.~Madore, P.~Manousselis and G.~Zoupanos, 
J. High Energy Phys. \andvol{04,2004,034}.

\bibitem{S4}
Y.~Azuma, S.~Bal, K.~Nagao and J.~Nishimura, 
J. High Energy Phys. \andvol{07,2004,066}.

\bibitem{YK-Gspin}
Y.~Kawamura, \PTP{110,2003,579}; hep-th/0304149.

\bibitem{YK-GHD}
Y.~Kawamura, \PTP{112,2004,299}; hep-th/0404044.

\bibitem{YK}
Y.~Kawamura, \PTP{107,2002,1105}; hep-th/0203007:
\andvol{109,2003,1}; hep-th/0206184:
\andvol{109,2003,153}; hep-th/0207054.

\bibitem{Nambu}
Y.~Nambu, \PR{D7,1973,2405}. 

\bibitem{p-brane}
E.~Bergshoeff, E.~Sezgin, Y.~Tanii and P.K.~Townsend, Ann. Phys. \andvol{199,1990,340}.

\bibitem{Filippov}
V.~A.~Filippov, Sib. Math. J. \andvol{26,1985,126}.
 
\bibitem{Hoppe}
J.~Hoppe, Helv. Phys. Acta. \andvol{70,1997,302}.

\bibitem{Xiong}
C.-S. Xiong, \PL{B486,2000,228}.

\bibitem{BFSS}
T.~Banks, W.~Fischler, S.~H.~Shenker and L.~Susskind, \PR{D55,1997,5112}.

\bibitem{Taylor}
I.~W.~Taylor, hep-th/0002016.

\bibitem{many-index}
H.~Awata, M.~Li, D.~Minic and T.~Yoneya, J. High Energy Phys. \andvol{02,2001,013}.

\bibitem{Mimic}
D.~Minic, hep-th/9909022.

\bibitem{IIB}
N.~Ishibashi, H.~Kawai, Y.~Kitazawa and A.~Tsuchiya, \NP{B498,1997,467}.

\bibitem{cohom}
R.~Bott and L.~Tu, {\it Differential forms in algebraic topology} (Springer, 1982).\\
C.~Nath and S.~Sen, {\it Topology and Geometry for Physicists} (Academic Press, London, 1983).

\bibitem{non-bos}
E.~Witten, \CMP{92,1984,455}.

\bibitem{brane-scan}
A.~Achucarro, J.~Evans, P.~Townsend and D.~Wiltshire, \PL{B198,1987,441}.

\bibitem{Schild}
A.~Schild, \PR{D16,1977,1722}.

\end{thebibliography}
\end{document}